\documentclass[12pt,twoside]{article}
\def\TheAuthor{Yuriy Mokrousov and Frank Freimuth}                                 
\def\TheTitle{Geometric Phases and Topological effects\footnotemark}          
\def\TheHeading{Geometric Phases and Topological effects}                                     
\def\TheInstitute{Peter Gr\"unberg Institut (PGI-1) and Institute for Advanced
Simulation (IAS-1)}   
\def\TheAddress{Forschungszentrum J\"ulich GmbH}         
\def\ThePart{A}                  
\def\TheChapter{6} 
\usepackage{ferienschule_14}
\usepackage{braket}
\usepackage{subfigure}
\usepackage{float}
\restylefloat{table}

\newcommand{\polla}{\mathbf{P}^{\lambda}}

\newcommand{\pollader}{\partial_{\lambda}\mathbf{P}^{\lambda}}
\newcommand{\depol}{\Delta\mathbf{P}}

\newcommand{\enla}{\varepsilon^{\lambda}_{n\mathbf{k}}}
\newcommand{\emla}{\varepsilon^{\lambda}_{m\mathbf{k}}}
\newcommand{\enk}{\varepsilon_{n\mathbf{k}}}

\newcommand{\intbz}{\int_{\rm BZ}d\mathbf{k}}
\newcommand{\unkl}{u^{\lambda}_{n\mathbf{k}}}
\newcommand{\umkl}{u^{\lambda}_{m\mathbf{k}}}

\newcommand{\pnkl}{\psi^{\lambda}_{n\mathbf{k}}}
\newcommand{\pmkl}{\psi^{\lambda}_{m\mathbf{k}}}

\newcommand{\nkay}{n\mathbf{k}}
\newcommand{\mkay}{m\mathbf{k}}
\newcommand{\bfk}{\mathbf{k}}
\newcommand{\bfq}{\mathbf{q}}
\newcommand{\bfE}{\mathbf{E}}
\newcommand{\bfeee}{\mathbf{e}_3}

\newcommand{\bfee}{\mathbf{e}_2}
\newcommand{\bfr}{\mathbf{r}}
\newcommand{\bfd}{\mathbf{d}}
\newcommand{\bfkappa}{\boldsymbol{\varkappa}}
\newcommand{\bfR}{\mathbf{R}}
\newcommand{\bfp}{\mathbf{p}}
\newcommand{\bfM}{\mathbf{M}}
\newcommand{\bfm}{\mathbf{m}}
\newcommand{\bfx}{\mathbf{x}}
\newcommand{\bfn}{\mathbf{n}}
\newcommand{\bfB}{\mathbf{B}}
\newcommand{\bfA}{\mathbf{A}}
\newcommand{\bfG}{\mathbf{G}}
\newcommand{\bfv}{\mathbf{v}}
\newcommand{\bfOmega}{\boldsymbol{\Omega}}
\newcommand{\bfsigma}{\boldsymbol{\sigma}}
\newcommand{\bfcalA}{\boldsymbol{\mathcal{A}}}
\newcommand{\calA}{\mathcal{A}}
\newcommand{\calC}{\mathcal{C}}

\makeindex
\begin{document}
\MakeTitel 
\footnotetext{Lecture Notes of the $45^{{\rm th}}$ IFF Spring
School ``Computing Solids - Models, ab initio methods and supercomputing''
(Forschungszentrum J{\"{u}}lich, 2014). All rights reserved. }
\tableofcontents
\newpage

\section{Introduction}

Since the beginning of the eighties, and especially since the seminal work by Berry~\cite{berry}, the mathematical
concepts of geometry, topology, geometric phase and topological characterization of so-called fibre bundles,
rapidly entered various  aspects of condensed matter physics such as, among others, electric polarization, 
Hall effect in insulators and metals, transport properties at surfaces and magnetization dynamics. Especially recently, triggered by
discovery of topological and spontaneous Chern insulators, the topological classification of solids based on 
quantities derived from their geometrical properties has become a very common tool in characterization of
physical properties of metals and insulators.  Moreover, interplay of complex magnetism with topological
properties of solids is being studied very intensively nowadays. 

It is impossible to review all of the issues listed above with a descent degree of depth within the format of
comparatively short lecture notes. We have thus decided to focus on selected aspects of the geometry-related
topics in condensed matter physics, trying to keep our manuscript as concise and as self-contained as 
possible. An interested reader should be able to follow our notes from the beginning to the end with minimal
reference to other sources. We start by discussing the mathematical foundations of the Berry, or, geometric
phase, formulate the adiabatic approximation for quantum dynamics and introduce fundamental concepts such
as Berry connection, Berry curvature, gauge freedom, parallel transport and first Chern number. The physical
examples we choose to apply the introduced concepts to are the Aharonov-Bohn effect and spin-$\frac{1}{2}$
in magnetic field, which prove to be of great importance for understanding the material in the rest of the notes.
We further show in detail the geometrical and topological nature of electric polarization in insulators, bringing 
to attention its relation to the Chern number. In a separate chapter we discuss, referring to simple arguments,
the emergence of the Chern insulators in two-dimensional reciprocal space, and the interplay between various
versions of, quantized and not, Hall effects taking place in metals and insualtors. We derive the expression for
the velocity of a state due to time-evolution of the quantum system and express it in geometrical terms, and
use it to arrive at the equations of motion which govern the dynamics of electrons in a solid in response to 
electro-magnetic fields and general perturbations. We discuss the consequences of these equations for such
properties as Hall conductance and orbital magnetization. Finally, as an example of a system with properties 
dependent on a parameter slowly varying in real space we choose a solid with spatially varying 
magnetization direction. We show how reformulation of the problem in geometrical terms can be used to
explain the rise of emergent magnetic field in such systems, and the emergence of the topological Hall effect.
We conclude with the purely geometric semiclassical derivation of the Dzyaloshinskii-Moriya interaction.

There is a number of books and reviews in which most of the aspects presented here can be found, and 
in these sources an interested reader can find more details and more profound discussions. Namely, we
refer to the books by Bohm~\cite{Bohm} and Nakahara~\cite{Nakahara} for a mathematically rigorous
discussion of geometric phase, fibre bundle theory and dynamics of quantum systems, briefly outlined
in section 2. The subject of electric polarization presented in section 3.2 has been discussed in depth 
in~\cite{resta-vanderbilt}. Various aspects of topological and Chern insulators are concisely and transparently
presented and discussed in~\cite{Bernevig}. The best review of the issues of Berry phase related electron
dynamics in external perturbations can be found in Ref.~\cite{Xiao}. A good introduction to the topological Hall effect
is given in~\cite{Murakami}. All corresponding original references can be found in  listed above works and we direct an interested reader there. Overall, we are also grateful to Hongbin Zhang, Robert Bamler, Achim Rosch, Gustav Bihlmayer
and Stefan Bl\"ugel for multiple discussions on the subject, scientific collaborations on selected aspects, and helping in determining the content of the manuscript and making its streaming more transparent. 

Finally, we remark that, for reasons of simplification, in the following we assume that $m_e=\hbar=e=c=1$. 
Most of the expressions which include the prefactors and which should be used for practical evaluation of the quantities,
can be found in the sources listed above. 

\section{General Theory}

\subsection{Berry phase and adiabatic evolution}
The origins of the Berry phase lie in the dynamics of a quantum system described by a Hamiltonian 
$H(\lambda)$ which bares a parametric dependence on a parameter $\lambda$. It is normally assumed that
$\lambda$ lies on a certain differentiable manifold $M$, $\lambda\in M$. We assume
that the Hamitonian $H(\lambda)$ as well as its descrete eigenspectrum $\{ \varepsilon_n(\lambda)\}$ are  
smooth and unique functions of $\lambda$ everywhere on $M$. For each $\lambda$ we denote by 
$\{\Ket{n\lambda}\in\mathcal{H}\}$ a set of ``instantaneous" solutions of 
\begin{equation}\label{equation1}
H(\lambda)\Ket{n\lambda}=\varepsilon_n(\lambda)\Ket{n\lambda}.
\end{equation}
It is important to realize that, in contrast to Hamiltonian and eigenvalues, each function $\Ket{n\lambda}:M\rightarrow\mathcal{H}$ can be chosen to be smooth only over certain parts of $M$ (called {\it patches}), but not necessarily over the
whole of $M$. Imagine two patches $O_1$ and $O_2$ in $M$, $O_1\cap O_2\neq\varnothing$, with two sets of 
smooth functions $\{\Ket{n\lambda}\}$ and $\{\Ket{n\lambda}'\}$ on them. Then for any $\lambda\in O_1\cap O_2$
we know that $\Ket{n\lambda}$ and $\Ket{n\lambda}'$ can differ only by a complex phase which is the element 
of group $\mathbb{U}(1)$:
\begin{equation}\label{gauge-trans}
\Ket{n\lambda}' = e^{i\zeta_n(\lambda)}\Ket{n\lambda}.
\end{equation}
Also on the patches themselves we can always switch from $\{\Ket{n\lambda}\}$ to the ``alternative" functions
$\{\Ket{n\lambda}'\}$ via the {\it gauge transformation}\iffindex{gauge transformation} given by (\ref{gauge-trans}) with arbitrary functions $\{ \zeta_n(\lambda)\}$ being smooth on corresponding patches. Indeed, both $\{\Ket{n\lambda}\}$ and $\{\Ket{n\lambda}'\}$ constitute
a possible set of instantaneous solutions of (\ref{equation1}) and the freedom in choice of either one or another
manifests the {\it gauge freedom}.\iffindex{gauge freedom} The corresponding group which is used to formulate the condition of the gauge
freedom is called the {\it gauge group}. In our case the gauge group is $\mathbb{U}(1)$.

The time evolution which we want to consider is realized via a certain time-dependence of 
$\lambda$, which goes along a given curve $\mathcal{C}$ in $M$: $t\in [0,T]\rightarrow \lambda(t)\in \mathcal{C}$. We will assume in the following
that $T$ is the period of $H(t)=H(\lambda(t))$,~i.e.~$\lambda(0)=\lambda(T)$, $H(0)=H(T)$, $\varepsilon_n(0)=\varepsilon_n(T)$. We will also assume for simplicity that closed path $\mathcal{C}$ completely lies in a single patch,~i.e.~we can choose $\Ket{n\lambda}$ smoothly and uniquely on $\mathcal{C}$, $ \Ket{n\lambda(0)}=\Ket{n\lambda(T)}$. We seek for solutions of the Schr\"odinger equation:
\begin{equation}\label{SCH}
i\frac{\partial\psi(t)}{\partial t}=H(\lambda(t))\psi(t),\quad \psi(t)\in\mathcal{H}.
\end{equation}
Let us also assume that at time $t=0$, $\psi(0)=\Ket{n\lambda(0)}$ for certain $n$. We are also particularly interested
in solutions of (\ref{SCH}) which are periodic in time in the sense that for a certain $\tau$
\begin{equation}
\Ket{\psi(0)}\Bra{\psi(0)}=\Ket{\psi(\tau)}\Bra{\psi(\tau)}.
\end{equation} 
Generally speaking, depending on the speed with which $\lambda$ is changed in time, $\tau$ can be arbitrary. Since
the problem of finding solutions of Schr\"odinger equation for arbitrary $\tau$ is very broad, we restrict ourselves to
the case of $\tau=T$.

Consider first the case when $[H(t),H(t')]=0$ everywhere on path $\mathcal{C}$. Constant in time Hamiltonian obviously 
belongs to this class. Using the spectral resolution of the
time-evolution operator in this case~\cite{Bohm}, we can show that
\begin{equation}\label{stationary}
\psi(t)=e^{-i\int_0^t \varepsilon_n(\tau)\,d\tau}\psi(0)=e^{-i\alpha_{\rm dyn}(t)}\psi(0) = e^{-i\alpha_{\rm dyn}(t)}\ket{n\lambda(0)},
\end{equation}
where with $\alpha_{\rm dyn}(t)$ we denoted the {\it dynamical phase}: $\alpha_{\rm dyn}(t)=\int_0^t \varepsilon_n(\tau)\,d\tau$. The solution $\psi(t)$ from above is obviously {\it stationary}:
\begin{equation}
W(t)=\Ket{\psi(t)}\Bra{\psi(t)} = \Ket{\psi(0)}\Bra{\psi(0)} = W(0)=\Lambda_n(0),
\end{equation}
where $\Lambda_n(\lambda)=\Ket{n\lambda}\Bra{n\lambda}$.

Next we introduce an important less stringent assumption which is called {\it adiabatic approximation}.\iffindex{adiabatic approximation}
Within this approximation we assume that:
\begin{equation}\label{ADIAB}
W(t) = \Lambda_n(t) \quad \Longleftrightarrow \quad \Ket{\psi(t)}=e^{-i\alpha_{\psi}(t)}\Ket{n\lambda(t)},
\quad t\in[0,T].
\end{equation}
Let us see how reasonable this assumption for {\it adiabatic time-evolution} is. For this we rewrite the general solution
of the Schr\"odinger equation in the following form:
\begin{equation}
\ket{\psi(t)} = \sum_n c_n(t) 
\ket{n\lambda(t)},
\end{equation}
and substitute it into (\ref{SCH}). This gives the following equations for the coefficients:
\begin{equation}\label{QUASI}
\partial_tc_n (t)= -i\varepsilon_n(t)c_n(t)-\sum_{m}c_m(t)\braket{n\lambda(t)|\partial_t|m\lambda(t)}.
\end{equation}
Thus, if we start with a certain $c_n(0)=1, c_m(0)=0, m\neq n$, the adiabatic assumption is approximately 
fulfilled when $\Braket{n\lambda(t)|\partial_t|m\lambda(t)}$ are small for $m\neq n$. One can use Eq.~(\ref{equation1})
to show that the latter matrix element can be expressed like this:
\begin{equation}\label{T-coeff}
\braket{n\lambda(t)|\partial_t|m\lambda(t)} = \frac{\bra{n\lambda(t)}\partial_t H(t)\ket{m\lambda(t)}}{\varepsilon_m(t)-\varepsilon_n(t)} = T^{-1}_{mn},
\end{equation}
which corresponds to the frequency of transition between the states $\ket{n\lambda(t)}$
and $\ket{m\lambda(t)}$. It is therefore for the states which are well-separated from each other
in energy and for slowly varying (slower than the intrinsic time-scale of quantum transitions between states) 
in time Hamiltonians that the adiabatic approximation is more valid. It is important to remember, however,
that the adiabatic wavefunction given by Eq.~(\ref{ADIAB}) can never be a solution of the Eq.~(\ref{SCH})
and can serve only as an approximation to it. 

Under adiabatic assumption we can directly solve Eq.~(\ref{QUASI}) for $c_n(t)$ by integration:
\begin{eqnarray}\label{adiabatic1}
c_n(t) & = &e^{-i\int_0^t\varepsilon_n(\tau)\,d\tau}\,e^{i\int_0^ti\braket{n\lambda(\tau)|\partial_{\tau}|n\lambda(\tau)}\,d\tau}=
e^{-i\alpha_{\rm dyn}}\,e^{i\gamma_n(t)},\\\label{adiabatic2}
\psi(t) & = &e^{-i\alpha_{\rm dyn}(t)}\,e^{i\gamma_n(t)}\Ket{n\lambda(t)},
\end{eqnarray}
where in addition to the dynamical phase the wavefunction acquires the so-called {\it geometric}, or, {\it Berry} phase factor
with the geometric phase $\gamma_n$:\iffindex{Berry phase}\iffindex{geometric phase}
\begin{equation}\label{GEO}
\gamma_n(t) = \int_0^ti\braket{n\lambda(\tau)|\partial_{\tau}|n\lambda(\tau)}\,d\tau\quad{\rm mod }\, 2\pi.
\end{equation}

\subsection{Connection and curvature}

The expression for the geometric phase (\ref{GEO}) can be easily recast into a time-independent purely geometrical representation:
\begin{equation}\label{BF}
\gamma_n(t) = \int_0^ti\braket{n\lambda(\tau)|\partial_{\tau}|n\lambda(\tau)}\,d\tau = \int_{\lambda(0)}^{\lambda(t)}
i\braket{n\lambda(\tau)|\partial_{\lambda_i}|n\lambda(\tau)}\,d\lambda_i = \int_{\lambda(0)}^{\lambda(t)}\mathcal{A}^n,
\end{equation}
where $\mathcal{A}^n$ is the so-called {\it Berry connection = connection = connection form}:\iffindex{Berry connection}\iffindex{gauge potential}
\begin{equation}\label{conn-A}
\mathcal{A}^n = i\braket{n\lambda|\partial_{\lambda_i}|n\lambda}\,d\lambda_i = \bfcalA^n\,d\boldsymbol{\lambda},\quad
\bfcalA^n_i=i\braket{n\lambda|\partial_{\lambda_i}|n\lambda}.
\end{equation}
It can be easily shown that $\calA^n$, and so is the geometric phase, are purely real quantities. Since function $\Ket{n\lambda}$ is smooth and unique on a given patch, $\calA^n$ is also a smooth and unique function on the corresponding
patch of $M$. At the boundary between two patches where $\Ket{n\lambda}$ and $\Ket{n\lambda}'$ are related by a 
gauge transformation (\ref{gauge-trans}), the corresponding connections are related via the gauge transformation:
\begin{equation}\label{gauge-A}
\calA'^n = \calA^n - d\zeta_n,\quad \bfcalA'^n = \bfcalA^n - \nabla\zeta_n.
\end{equation}
Also upon a change in the gauge of the $\Ket{n\lambda}$ on the patch itself the connection transforms according to (\ref{gauge-A}). Mathematically~\cite{Nakahara}, the family of patches $\{O\}$ from $M$, together with the corresponding
family of connections $\{\calA^n\}$ and $\mathbb{U}_1$-gauge transformations between the patches $\{ \zeta_n \}$ endows $M$ with the geometric structure suitable for studies of many problems in quantum physics. The mathematical name of this structure is a $\mathbb{U}_1$ {\it principle fibre bundle}. The quantum theory formulated in terms of a principle fibre bundle theory with a certain group $\mathbb{G}$ of gauge transformations is often called a $\mathbb{G}$ {\it gauge theory}. For example, theory of electromagnetism can be elegantly recast in terms of the fibre bundle theory, with the role of connection played by the electro-magnetic vector potential $\bfA$~\cite{Nakahara}. As we shall see, the analogy between the Berry phase theory and electromagnetism goes much deeper than sharing the name for corresponding geometrical structure. 

We now consider a closed path $\mathcal{C}$ with $\lambda(0)=\lambda(T)$ and $\gamma_n(\mathcal{C}):=\gamma_n(T)$.
Let us trace how $\gamma_n(\mathcal{C})$ changes upon an arbitrary smooth and unique change of gauge in the patch in 
which $\calC$ lies:
\begin{equation}
\gamma'_n(\calC)=i\oint_{\calC}\calA'^n=i\oint_{\calC}\calA^n-i\oint_{\calC}d\zeta_n = i\oint_{\calC}\calA^n=\gamma_n(\calC),
\end{equation}
since $\zeta_n(0)=\zeta_n(T)$. This means that the Berry phase of closed path $\calC$ is an intrinsically {\it gauge-invariant}
property,\iffindex{gauge invariance} which is another manifestation of its purely geometrical meaning. One can show that $\gamma_n(\calC)$ remains
constant even under action of a more general class of gauge transformation, and cannot be ``gauged away". While it cannot
be done for the whole path, on a part of $\calC$ the Berry phase can be indeed gauged away by choosing the so-called
{\it parallel transport gauge} realized by smooth and unique functions $\zeta^{\rm PT}_n$, defined only on the part of 
$\calC$ which we denote as $\tilde{\mathcal{C}}$. Given the initial choice of $\{ \Ket{n\lambda}\}$, $\zeta^{\rm PT}_n$ 
functions satisfy the following condition:
\begin{equation}
\bfcalA^{\rm PT,n}=\bfcalA^n - \nabla\zeta^{\rm PT}_n=0,  
\end{equation}
equivalent to the condition that 
\begin{equation}\label{PT-nl}
\braket{n\lambda^{\rm PT}|\nabla_{\lambda}|n\lambda^{\rm PT}}=0, \quad \gamma^{\rm PT}_n(\tilde{\calC}):=i\int_{\tilde{\calC}}\calA^{{\rm PT},n}=0.
\end{equation}
Besides being very convenient, the parallel transport gauge is also easy to constuct. Namely, given a certain starting point
$\lambda$ and a set of starting $\{\Ket{n\lambda}^{\rm PT}\}$ at this point, a set of ``parallel transported" instantaneous states at an infinitesimally to $\lambda$ close point $\lambda+\epsilon$ can be constucted using perturbation theory:
\begin{equation}\label{PT-gauge}
\Ket{n,\lambda+\epsilon}^{\rm PT} = \Ket{n\lambda}^{\rm PT} + \sum_{m\neq n}\frac{\braket{m\lambda^{\rm PT}|H(\lambda+\epsilon)-H(\lambda)|n\lambda^{\rm PT}}}{\varepsilon_n(\lambda)-\varepsilon_m(\lambda)}\Ket{m\lambda^{\rm PT}}.
\end{equation}
It can be easily checked that such constructed instantaneous solutions are indeed ``parallel transported". It is common to 
use the parallel transport gauge for evaluation of the quantities which are gauge-invariant {\it locally} for each point $\lambda$ on $M$. A most important example of such quantity is the (Berry) curvature. 

{\it (Berry) curvature} is defined as a rank-2 antisymmetric tensor with the components given by:\iffindex{Berry curvature}
\begin{equation}\label{Berry-C}
\Omega_{ij}^n:=\partial_{\lambda_i}\calA^n_j- \partial_{\lambda_j}\calA^n_i=-2{\rm Im}\Braket{\frac{\partial}{\partial_{\lambda_i}}n\lambda|\frac{\partial}{\partial_{\lambda_j}}n\lambda}.
\end{equation}
It is easy to see that the components of $\Omega^n$ are {\it locally} gauge-invariant with respect to the gauge transformations
(\ref{gauge-A}). In the language of differential forms, $\Omega^n$ is a 2-form given by $\Omega^n=d\calA^n$.
Using the Stokes' theorem the expression for the Berry phase (\ref{BF}) can be rewritten as:
\begin{equation}
\gamma_n(\calC)= i\oint_{\calC}\calA^n = \int_S \Omega^n,
\end{equation}
where $S$ is a ``surface" in $M$ which $\calC$ encompasses: $\partial S=\calC$. The Berry phase thus equals to the 
{\it flux} of the Berry curvature through $S$. A very useful expression
for the curvature can be obtained within the parallel transport gauge~(\ref{PT-gauge}), in which the perturbation theory
expression for the derivative of the $\Ket{n\lambda}$ entering (\ref{Berry-C}) can be written down (we omit the "PT"
label for simplicity):
\begin{equation}\label{pert-theo}
\partial_{\lambda_i}\Ket{n\lambda} = \sum_{m\neq n}\frac{\braket{m\lambda|\partial_{\lambda_i}H(\lambda)|n\lambda}}{\varepsilon_n(\lambda)-\varepsilon_m(\lambda)}\Ket{m\lambda},
\end{equation}
which gives the following gauge-invariant expression for the curvature:
\begin{equation}\label{l-curv}
\Omega_{ij}^n(\lambda) = -2{\rm Im}\sum_{m\neq n}\frac{\braket{m\lambda|\partial_{\lambda_i}H(\lambda)|n\lambda}\braket{n\lambda|\partial_{\lambda_j}H(\lambda)|m\lambda}}{[\varepsilon_n(\lambda)-\varepsilon_m(\lambda)]^2}.
\end{equation}
The latter expression is not only useful practically, but it is also rather instructive since it underlines the role 
of the band degeneracies as the sources of the Berry curvature around them. We will see several examples of
connection between the band degeneracies and the curvature in the course of this manuscript.

In case of $M$ being a part of $\mathbb{R}^3$ the curvature tensor can be seen as a vector $\bfOmega^n$ in $\mathbb{R}^3$, with components given by:
\begin{equation}
\bfOmega_i^n:=(1/2)\epsilon_{ijk}\Omega^n_{jk},
\end{equation} 
and the relation (\ref{Berry-C}) between the connection and curvature becomes:
\begin{equation}
\bfOmega^n={\rm curl}\bfcalA^n.
\end{equation} 
There can be cases  when the connection is the so-called ``pure gauge" connection,~i.e., there exists a smooth and unique vector field $\mathbf{f}$ on $M$, so that $\bfcalA^n=\nabla \mathbf{f}$. Clearly, in this case the Berry curvature equals zero identically on $M$ and the Berry phase is vanishing for any given 
closed path in $M$. In the rest of the manuscript we are particularly interested in situations for which this does not happen.

Finally, we would like to remark on the occurrence of the so-called {\it non-Abelian} Berry phase. It differs from the 
{\it Abelian} case considered above in that one has to deal with an $N$-fold degeneracy of  eigenvalue $\varepsilon_n$
at any point $\lambda$  from $M$. Given a certain set of instantaneous solutions $\{\Ket{n\alpha\lambda}\}_{\alpha=1}^N$ which correspond to $\varepsilon_n$ for each $\lambda$ we can construct corresponding eigenprojectors $\Lambda_n(\lambda)=\sum_{\alpha=1}^N\Ket{n\alpha\lambda}\Bra{n\alpha\lambda}$. The eigenprojectors are constant with respect to the gauge transformations realized by the unitary transformations $\mathcal{U}^N$ from $\mathbb{U}(N)$: 
\begin{equation}\label{U-gauge}
\vec{\Ket{n\alpha\lambda}}' = \mathcal{U}^N\vec{\Ket{n\alpha\lambda}},
\end{equation}
where $\vec{\Ket{n\alpha\lambda}}$ is a vector consisting of ordered instantaneous solutions. The adiabatic assumption for the non-Abelian case can be formulated in analogy to the Abelian case: the solution of the Schr\"odinger equation has to 
reside in $n$'th eigenspace, $\Ket{\psi(t)}\Bra{\psi(t)}=\Lambda_n(\lambda(t))$. The most general shape of the wavefunction which solves (\ref{SCH}) under the adiabatic assumption reads:
\begin{equation}
\Ket{\psi(t)}=\sum_\alpha c_{\alpha}^n(t)\Ket{n\alpha \lambda(t)}.
\end{equation}
Substituting the latter expression into (\ref{SCH}) gives us a system of equations for the $c$-coefficients, which
can be solved in analogy to the Abelian case yielding:
\begin{equation}\label{c-non-Ab}
\Vec{c^n}(t)=\mathcal{T}\exp \left(  \int_0^t d\tau\left\{  -i\varepsilon_n\mathbb{I} +
i\calA_{N}^n(\tau) \right\} \right)\vec{c^n}(0),
\end{equation}
where $\mathcal{T}$ is the time-ordering operator, $\mathbb{I}$ is the $N\times N$ identity matrix, and $\calA_{N}^n$ 
is a Hermitian-matrix valued 1-form, given by components:
\begin{equation}
\left[ \calA_{N}^n(\lambda(t))\right]^{\alpha\beta} := i\braket{n\alpha\lambda(t)|\partial_t|n\beta\lambda(t)}dt.
\end{equation}
The form $\calA_{N}^n$ is called the {\it non-Abelian (Berry) connection}, in analogy to the Abelian connection. The 
gauge transformation of the non-Abelian connection can be deduced from (\ref{U-gauge}):
\begin{equation}\label{non-Ab-A}
\calA^{\prime n}_N = (\mathcal{U}^N)^{-1}\calA^{n}_N\,\mathcal{U}^N +i(\mathcal{U}^N)^{-1}d\,\mathcal{U}^N.
\end{equation}
Obviously this relation reduces to (\ref{gauge-A}) if $N=1$. Equation (\ref{c-non-Ab}) allows us to calculate the 
total ``phase" a wavefunction acquires during adiabatic evolution with the period $T$ along a closed path $\calC$,~i.e., $\psi(T)=\mathcal{U}_{\psi}\psi(0)$, with the ``phase" $\mathcal{U}_{\psi}$ given by:
\begin{equation}
\mathcal{U}_{\psi} = \exp\left( -i\int_0^T\varepsilon_n(\lambda(t))dt \right)\,\mathcal{P}\exp\left(i\oint_\calC
\calA_N^n
\right),
\end{equation}
where the first factor is the analogon of the Abelian dynamical phase, while the second matrix is the {\it non-Abelian
Berry phase}. While having most of the geometric ``niceties" of the Abelian Berry phase, the non-Abelian phase is a
matrix with elements which are not separately gauge-invariant. Most relevant are thus the trace (also known as the 
Wilson loop) and the eigenvalues of the non-Abelian Berry phase, which are indeed gauge-invariant quantities. The 
expression for the matrix valued non-Abelian Berry curvature reads:
\begin{equation} \label{NAB}
\Omega_{N,ij}^n:=\partial_{\lambda_i}\calA^n_{N,j}- \partial_{\lambda_j}\calA^n_{N,i} + [\calA^n_{N,i},\calA^n_{N,j}],
\end{equation}
which reduces to (\ref{Berry-C}) for $N=1$.

\subsection{Topological phase: Aharonov-Bohm effect}

The Berry phase which we introduced is {\it geometric} in nature,~i.e., it depends
on the local geometry of the path $\mathcal{C}$. This also means that when the path $\mathcal{C}$ is changed, the  geometric phase will change as well. There can be
situations, however, when the Berry phase is not modified when the path $\mathcal{C}$ is subject to smooth transformations, in which case the Berry phase is {\it topological} in nature. A very prominent example of a topological phase arises within the {\it Aharonov-Bohm effect} (AB-effect).\iffindex{Aharonov-Bohm effect}

Imagine an infinitesimally small cylinder $D$ with the axis along $z$ and cutting through the center of coordinates 
in $\mathbb{R}^3$. Inside the cylinder only we generate a constant magnetic field $\bfB$ along the cylinder axis and 
a magnetic flux through the cylinder's cross section of $\Phi$. Consider an electron confined to a box which is positioned
at point $\bfR$ which lies very far away from the cylinder. The Hamiltonian of our electron with respect to the center of the box reads:
\begin{equation}
H=H(\bfp - \bfA(\bfx),\bfx-\bfR),
\end{equation}
where $\bfA$ is the electromagnetic vector potential with ${\rm curl} \bfA=\bfB$. We can thus look at the dependence
of the Hamitonian on $\bfR$ as a parametric dependence considered previously with $\lambda=\bfR$. According to the theory above, we have
to find the eigenvalues and eigenvectors of $H$:
\begin{equation}
H(\bfp - \bfA(\bfx),\bfx-\bfR)\Ket{n\bfR}=\varepsilon_n(\bfR)\Ket{n\bfR}.
\end{equation}
It is easy to see that to find $\{ \Ket{n\bfR} \}$ we can use the eigenvalues $\varepsilon_n(\bfR)$ and eigenfunctions 
$\psi^0_n$ of the free Hamiltonian
$H_0=H(\bfp,\bfx-\bfR)$:
\begin{equation}\label{AB-section}
\Braket{\bfx|n\bfR} = e^{i\int_{\bfR}^\bfx \bfA\, d\mathbf{l}}\,\psi_n^0(\bfx-\bfR),
\end{equation}
where the path between $\bfx$ and $\bfR$ cannot pass through $D$. We can thus readily show that the Berry
connection $\bfcalA_n$ is given by:
\begin{eqnarray*}
\bfcalA^n = i\Braket{n\bfR|\nabla_{\bfR}|n\bfR} &=& i\int_{\mathbb{R}^3}d\bfx\,
 \psi_n^{0*}(\bfx-\bfR)\left\{-i\bfA(\bfR)\psi_n^0(\bfx-\bfR) + \nabla_\bfR\psi_n^0(\bfx-\bfR)
\right\}\\
&=& \bfA.
\end{eqnarray*}
In turn the curvature is given by:
\begin{equation}\label{232}
\bfOmega^n = {\rm curl} \bfcalA={\rm curl} \bfA= \bfB.
\end{equation}
This means that for our problem the Berry phase $\gamma_n(\mathcal{C})$ acquired during adiabatic motion of our electron 
box around the cylinder with the magnetic field along a path $\mathcal{C}$ is given by:
\begin{equation}
\gamma_n (\mathcal{C})=\oint_{\mathcal{C}} \bfA\,d\mathbf{l} = -\Phi.
\end{equation}
Note that $\gamma_n$ does not depend neither on the ``band" index $n$ nor on the path $\mathcal{C}$, as long as
the cylinder is encircled the same number of times. This presents an elegant example of the topological phase which
does not change upon smooth transformations of $\mathcal{C}$. What is very important for us is that the results of
Eq.~(\ref{232}) can
be generalized to the case of electrons not necessarily confined to a box, with the magnetic field $\bfB$ present 
everywhere in $\mathbb{R}^3$ and possibly having non-trivial spatial distribution. This will result in the Berry phase
restoring back its geometrical rather than topological meaning, since in the latter case:
\begin{equation}
\gamma_n (\mathcal{C})=\oint_{\mathcal{C}} \bfA\,d\mathbf{l} = -\Phi(\mathcal{C}),
\end{equation}
with $\Phi(\mathcal{C})$ as the $\mathcal{C}$-dependent flux of the magnetic field through the area enclosed by 
$\mathcal{C}$.

\subsection{Dirac monopole and spin-$\frac{1}{2}$ problem}

As an example of the adiabatic dynamics we want to study the Hamiltonian of a spin-$\frac{1}{2}$
particle in magnetic field $\bfB=B\cdot\bfn$ in $\mathbb{R}^3$, with the unit vector $\bfn$ along $\bfB$, described by the Hamiltonian:
\begin{equation}\label{Mag-F}
H(\bfn)= b_0\bfsigma\cdot\bfB=b\bfsigma\cdot\bfn,
\end{equation}
where $b=b_0B$ is a constant which we would like to keep unspecified, and $\bfsigma$ is the vector of Pauli matrices. The parameter space of our problem is the 2-sphere $S^2$ embedded in $\mathbb{R}^3$,
with the parameter $\bfn$ playing the role of $\lambda$. We parametrize $\bfn$ in spherical coordinates: $\bfn=(\sin\theta\cos\varphi,\sin\theta\cos\varphi,\cos\theta)$, where $\theta$ and
$\varphi$ run between $0$ and $\pi$, and $0$ and $2\pi$, respectively, see Fig.~\ref{Sphere}. For each of the 
values of $\bfn$ we have to find the instantaneous eigenstates and eigenvalues of $b\bfsigma\cdot\bfn$.
There are two eigenvalues for each $\bfn$: $\varepsilon_+=+b$ and $\varepsilon_-=-b$, and two
corresponding eigenstates of the projection of $\bfsigma$ onto $\bfn$: $\ket{+\bfn}$ and $\ket{-\bfn}$, or, alternatively:
\begin{equation}
H(\bfn)\ket{\sigma\bfn}=b\sigma\ket{\sigma\bfn},
\end{equation}
where $\sigma$ takes values of $+/-$ or $+1/-1$ depending on the situation.
Note that the eigenvalues do not depend on $\bfn$. More generally, the operator
$\bfsigma$ in (\ref{Mag-F}) can be replaced with $2\mathbf{J}$, where $\mathbf{J}$ is the
angular momentum operator of our quantum system. In this case $\sigma$ is
not anymore restricted to take only two values, and the problem can be reduced
to more than two copies of the problem for spin-up and spin-down states only.
Nevertheless the conclusions and derived expressions remain basically the same
as those for the spin-$\frac{1}{2}$ problem, which we are to consider below.

\begin{figure}[ht!]
\begin{center}
\includegraphics[width=0.52\textwidth]{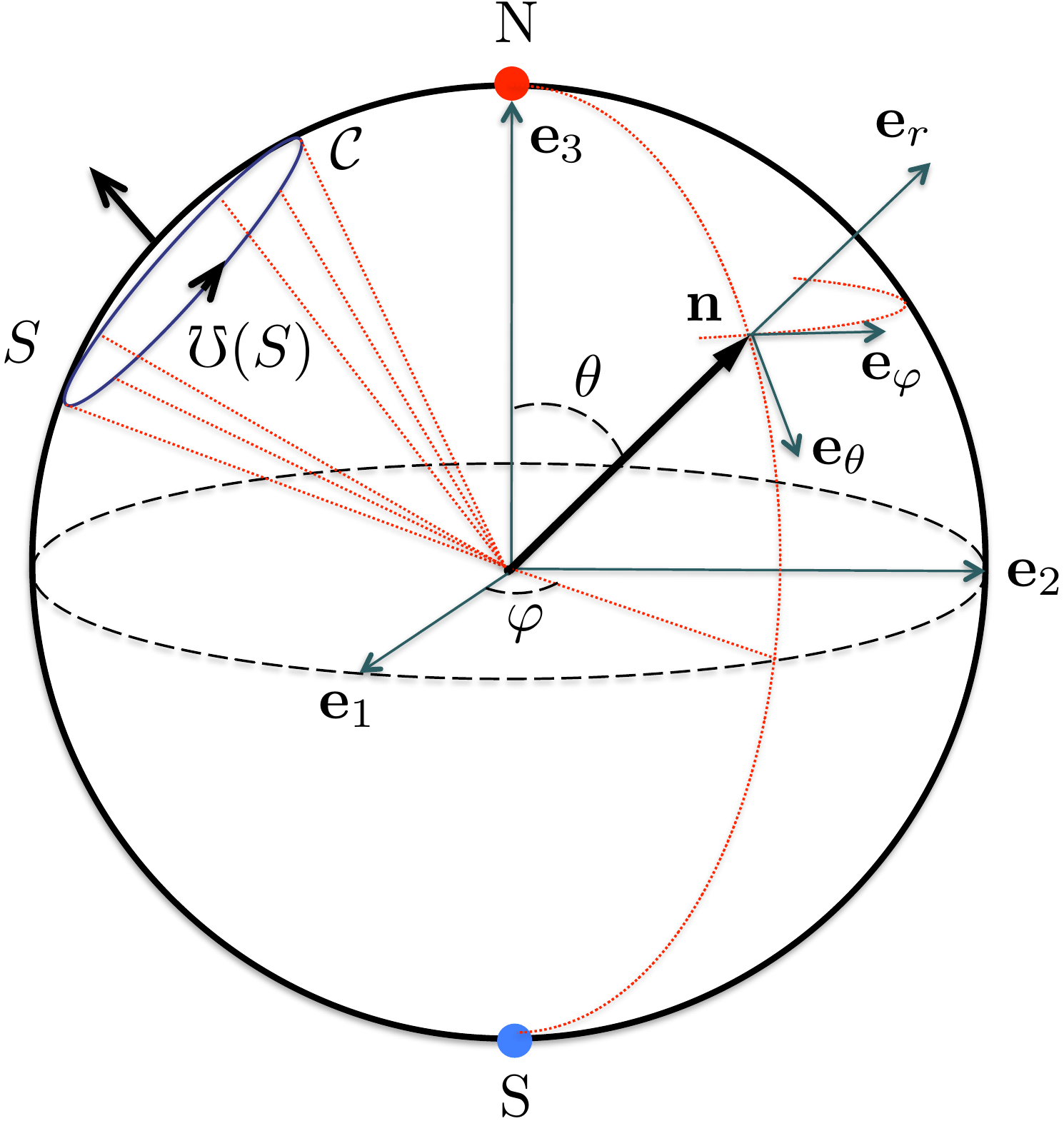}\vspace{1.0cm}
\includegraphics[width=0.7\textwidth]{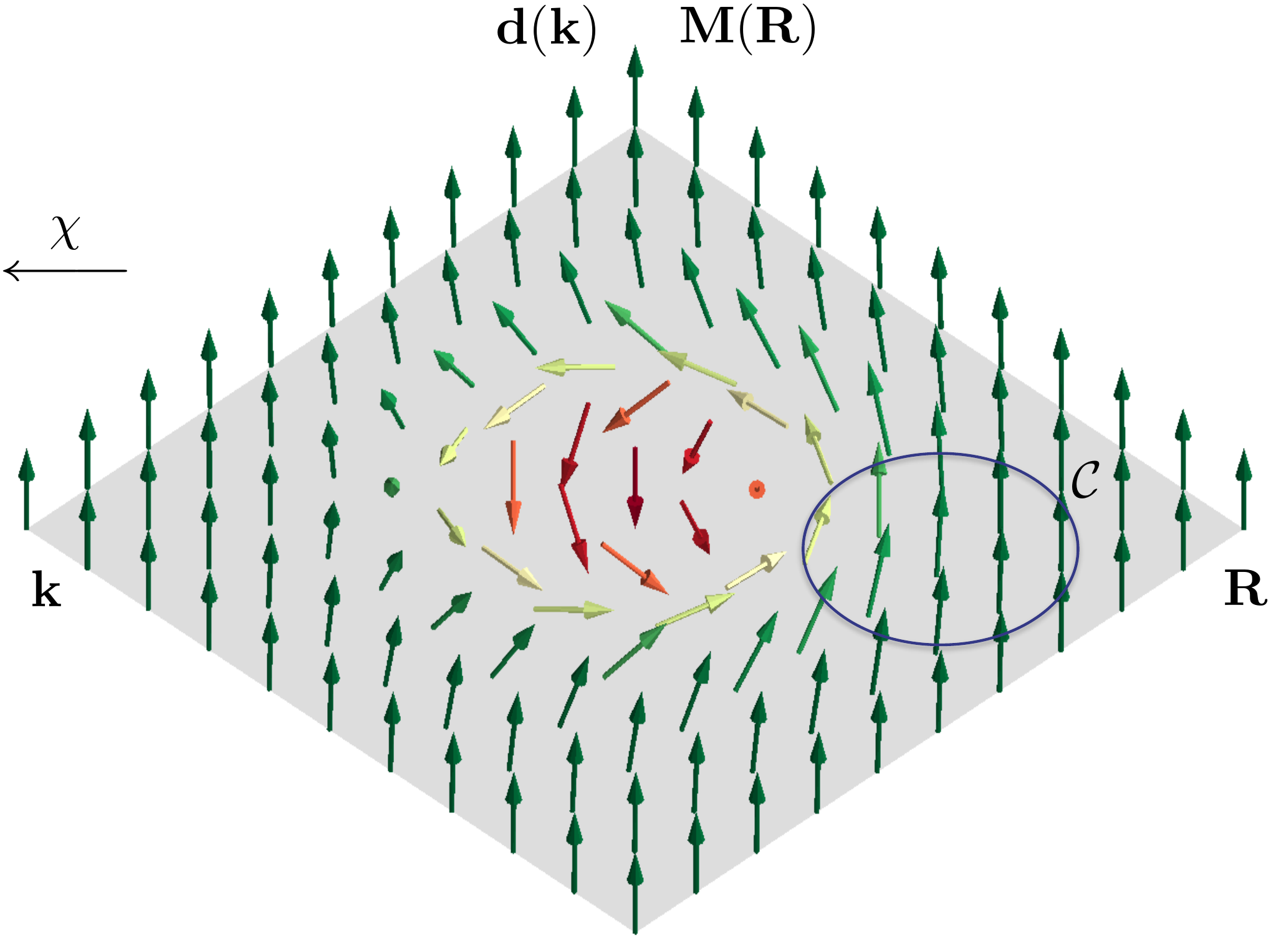}
\end{center}
\caption{\label{Sphere} Top: geometry of spin-$\frac{1}{2}$ in magnetic field. Bottom: skyrmion distribution of 
vector field $\bfd(\bfk)$ as a function of the Bloch vector $\bfk$ in Eq.~(\ref{111}); or of the magnetization field $\bfM(\bfR)$ as
a function of real-space coordinate $\bfR$ for magnetic texture as given by Eq.~(\ref{3.4-1}). The transformation
from the $\bfk(\bfR)$-space into the space of the magnetic field from spin-$\frac{1}{2}$ problem is realized by a map $\chi$,
which allows to relate the Berry phase of path $\calC$ in the lower figure to the Berry phase of the corresponding path $\calC$
in the upper figure.
The direction of $\bfd(\bfR)$ in the lower plot spans the whole sphere in the upper plot once. The artist's
view of the skyrmion is by J\"urgen Weischenberg.}
\end{figure}

Our task now is to choose as smooth a gauge of 
$\ket{+\bfn}$ and $\ket{-\bfn}$ as we can. We note that the vector $\bfn$ can be obtained from $\bfeee$ via the following 
${\rm S}\mathbb{O}(3)$ rotation: $\mathcal{R}(\theta,\phi)=\mathcal{R}_3(\varphi)\mathcal{R}_2(\theta)\mathcal{R}_3(-\varphi)$, where $\mathcal{R}_3(\varphi)$ performs rotation around $\bfeee$
by angle $\varphi$, and $\mathcal{R}_2(\theta)$ is a rotation around $\bfee$ by $\theta$. The corresponding rotation of the spinor wavefunction is an element of ${\rm S}\mathbb{U}(2)$ and is given by:
\begin{equation}\label{U-trans}
U(\theta,\varphi)=U_3(\varphi)U_2(\theta)U_3(-\varphi)=e^{-i\varphi\sigma_z/2}\,
e^{-i\theta\sigma_y/2}\,e^{i\varphi\sigma_z/2}.
\end{equation}
We thus define:
\begin{equation}\label{e3}
\ket{+\bfn} = U(\theta,\varphi)\ket{+\bfeee}, \quad \ket{-\bfn} = U(\theta,\varphi)\ket{-\bfeee}\quad\Leftrightarrow\quad \ket{\sigma\bfn} = U(\theta,\varphi)\ket{\sigma\bfeee}.
\end{equation}
It can be easily shown that such choice for $\ket{+\bfn}$ and $\ket{-\bfn}$ is smooth everywhere 
on $S^2$ except for the south pole S corresponding to vector $-\bfeee$. To achieve smoothness
at S, we need to perform a gauge transformation into a new basis:
\begin{equation}
\ket{\sigma\bfn}'=\ket{\sigma\bfn'}=e^{-i\sigma\varphi}\ket{\sigma\bfn},\quad\zeta_{\sigma}(\bfn) = -\sigma\varphi
\end{equation}
and it can be shown that the primed instantaneous eigenstates are smooth everywhere on $S^2$
except for the north pole N which corresponds to $\bfeee$. It is a property of the spin-$\frac{1}{2}$ system
in magnetic field that enforces us to introduce at least two patches on $S^2$ where the instantaneous eigenstates are smooth, and a phase ``twist" $=$ gauge transformation between two smooth families
at the overlap between the two patches. We denote these patches by $O_1=S^2-{\rm S}$ and $O_2=S^2-{\rm N}$. Note that we could also freely choose for $O_1$ and $O_2$ the northern and southern hemispheres including the equator. In this case the gauge transformation would act only at the equator
(to be rigorous, since the patches have to be open sets, by equator here we mean an infinitesimally small
open ``belt" which includes the equator).

We are now ready to define the connections on $O_1$ and $O_2$. They are given by corresponding
components:
\begin{eqnarray}
\calA^{\sigma}_{\theta} (\theta,\varphi)= \Braket{\sigma\bfn|\partial_{\theta}|\sigma\bfn}=0,\quad 
\calA^{\sigma}_{\varphi} (\theta,\varphi)= \Braket{\sigma\bfn|\partial_{\varphi}|\sigma\bfn}=-(1/2)\sigma(1-\cos\theta),\\
\calA'^{\sigma}_{\theta} (\theta,\varphi)= \Braket{\sigma\bfn'|\partial_{\theta}|\sigma\bfn'}=0,\quad 
\calA'^{\sigma}_{\varphi} (\theta,\varphi)= \Braket{\sigma\bfn'|\partial_{\varphi}|\sigma\bfn'}=(1/2)\sigma(1+\cos\theta).
\end{eqnarray}
Indeed, according to the gauge transformation rule, $\calA^{\sigma}_{\varphi} - \calA'^{\sigma}_{\varphi}=-\sigma=\partial_{\varphi}\zeta_{\sigma}$. The only component of the curvature can be also readily evaluated:
\begin{equation}
\Omega^{\sigma}_{\theta\varphi}(\theta,\varphi)=-\sigma\sin\theta/2.
\end{equation}
Since our sphere $S^2$ is naturally embedded into $\mathbb{R}^3$, the connection and curvature 
can be recast as vectors in $\mathbb{R}^3$:
\begin{equation}
\bfcalA^{\sigma} = \hat{\mathcal{A}}^{\sigma}_r \mathbf{e}_r + \hat{\mathcal{A}}^{\sigma}_\theta \mathbf{e}_\theta + \hat{\mathcal{A}}^{\sigma}_\varphi \mathbf{e}_\varphi = \hat{\mathcal{A}}^{\sigma}_\varphi \mathbf{e}_\varphi,\quad
\bfcalA'^{\sigma} = \hat{\mathcal{A}}'^{\sigma}_\varphi \mathbf{e}_\varphi,\quad
\bfOmega^{\sigma} = \Omega_r^{\sigma}\,\mathbf{e}_r,
\end{equation}
where the components read:
\begin{equation}\label{Sp-A}
\hat{\mathcal{A}}^{\sigma}_\varphi = \frac{\sigma(\cos\theta-1)}{2r\sin\theta}, \quad 
\hat{\mathcal{A}}'^{\sigma}_\varphi = \frac{\sigma(\cos\theta+1)}{2r\sin\theta}, \quad
\Omega_r^{\sigma} = -\frac{\sigma}{2r^2}.
\end{equation}
The latter expressions are remarkable in that they present a realization of so-called {\it Dirac monopole}.\iffindex{Dirac monopole}
Dirac monopole corresponds to a problem of a single {\it monopole} magnetic charge of magnitude $g$ at the origin which provides the source of the magnetic field $\bfB_m$:
\begin{equation}
{\rm div} \bfB_m = 4\pi g \,\delta^3(\bfr),
\end{equation}
with the solution of this equation being
\begin{equation}
\bfB_m = \frac{g}{r^2}\,\mathbf{e}_r,
\end{equation}
which is exactly the expression for the curvature with $g=-\sigma/2$. The flux of this magnetic field through $S^2$ is $4\pi g$. Obviously, the magnetic field $\bfB_m$ and $\bfOmega^{\sigma}$ have a
singularity at $r=0$. As discussed previously, this singularity in the Berry curvature is due to a 
degeneracy between $\varepsilon_+$ and $\varepsilon_-$ at the origin. In this sense, we can say
that such degeneracy points in the spectrum serve as ``sources" of the Berry curvature everywhere
around them, in the same way that the magnetic monopole $g$ serves as a source of the magnetic 
field $\bfB_m$. It is important to remember that it is the Hamiltonian which we started with which
serves as a source of non-trivial geometry of our problem,~i.e., the fact that $\sigma\neq 0$ gives
rise to the non-trivial gauge transformation which in turn leads to the non-zero monopole charge
at the origin. However, as we shall see, the geometry of the problem can also have important 
consequences for the Hamiltonian and we will show that $\sigma$ {\it has to} be quantized, even though
such assumption was not necessary to make from the beginning. 

To show this, lets evaluate the Berry phase of a certain path $\mathcal{C}$ which our spin ``draws"
on $S^2$ as it follows the adiabatically slow magnetic field. If we suppose that 
$\mathcal{C}$ does not include the south pole, then we can easily write the Berry phase
as:
\begin{equation}
\gamma_{\sigma}(\mathcal{C})=\oint_{\mathcal{C}}\bfcalA^{\sigma}\,d\mathbf{l}=
\int_{S}\bfOmega^{\sigma}\,d\mathbf{S}= -\sigma\mho(S)/2\quad {\rm mod}\,2\pi,
\end{equation}
where $\mathcal{C}=\partial S$, $\mho(S)$ is the solid angle of surface $S$, and the orientations of $d\mathbf{l}$ and $d\mathbf{S}$ are shown in Fig.~\ref{Sphere}. On the other
hand, we can use surface $S'=S^2-S$ to perform the integral of the curvature:
\begin{equation}
\gamma_{\sigma}(\mathcal{C})= -\sigma\mho(S)/2  = -\int_{S'}\bfOmega^{\sigma}\,d\mathbf{S} = \sigma\mho(S')/2 = \sigma(4\pi - \mho(S))/2\quad {\rm mod}\,2\pi,
\end{equation}
and thus we independently arrive at the {\it quantization} of $\sigma$:
\begin{equation}
\sigma \in \mathbb{Z},
\end{equation}
which is the sole property of the geometry of our problem. This quantization condition leads us to the definition of the {\it (first) Chern number}:\iffindex{Chern number}
\begin{equation}
C_{\sigma} = \frac{1}{2\pi}\int_{S^2}\bfOmega^{\sigma}\,d\mathbf{S} = -\sigma,
\end{equation}
which stands for the flux of the Berry curvature through the whole sphere.
It can be shown that for ``well-behaved" two-dimensional compact manifolds for which we will also
assume that they have no boundary (e.g. sphere, torus)
the Chern number is a topological invariant of the manifold with the geometry 
on it specified by a family of connections $\{ \calA\}$ and gauge transformations
$\{ \zeta \}$ belonging to the gauge group (i.e.~of the so-called mathematically {\it fibre bundle}), 
and that $C \in \mathbb{Z}$. Intuitively, it is clear why the manifold has to have no boundary, i.e. to be closed. 
Imagine that instead of $S^2$ we have only a part of it. Then we can easily imagine how by a smooth deformation
we could ``pull out" the Dirac monopole out of the sphere, thus continuously changing the flux of the curvature
through the part of $S^2$. On the other hand, smooth transformations can only change the position of the monopole inside the
whole sphere while keeping the value of the flux quantized.
Our example
of spin-$\frac{1}{2}$ particle in magnetic field and proof of quantization of $C_\sigma$ is a particular example of this general mathematical fact. As we have also seen, the non-zeroness of the Chern number is intrinsically related to our inability
of choosing a smooth gauge over the {\it entire} manifold. We refer to \cite{Nakahara}
for further discussions on higher-dimensional manifolds, theory of Chern numbers and characteristic classes.

\section{Selected Applications of Geometric Phase in Condensed Matter}

\subsection{Berry curvature for Bloch electrons}

For electrons in a periodic solid the eigenstates $\psi_{n\mathbf{k}}$ of the 
Hamiltonian can be classified by quantum numbers $(\mathbf{k},n)$, 
where $\mathbf{k}$ lies in the so-called Brillouin
zone (BZ), and $n$ is a discrete index numbering the bands. The eigenfunctions
can be written in the following form:
\begin{equation}
\psi_{n\mathbf{k}}(\mathbf{r})=e^{i\mathbf{k\cdot r}}u_{n{\mathbf{k}}}(\mathbf{r}),
\end{equation}
where $u_{n\mathbf{k}}$ has the periodicity of the lattice. It then follows that
instead of writing 
\begin{equation}\label{Eq1}
H\psi_{n\mathbf{k}}(\mathbf{r})=\enk\psi_{n\mathbf{k}}(\mathbf{r})
\end{equation} 
we can write 
\begin{equation}
H(\mathbf{k})u_{n\mathbf{k}}(\mathbf{r})=\enk u_{n\mathbf{k}}(\mathbf{r}),
\end{equation} 
where 
\begin{equation}
H(\mathbf{k})\equiv H_{\bfk}\equiv e^{-i\mathbf{k\cdot r}}\cdot H\cdot e^{i\mathbf{k\cdot r}}.
\end{equation}
We were thus able to rewrite the problem~(\ref{Eq1}) in terms of an eigenvalue
problem of a $\mathbf{k}$-dependent Hamiltonian, acting on the same Hilbert
space of periodic functions for every $\mathbf{k}$. This is exactly the setup suitable
for studies of the Berry phase effects, if we identify the parameter $\lambda$ from general
mathematical theory with the Bloch vector $\mathbf{k}$. The corresponding (Berry) connection 
of non-degenerate band $n$ according to (\ref{conn-A}) then reads:
\begin{equation}
\mathcal{A}^n(\mathbf{k})=i\Braket{u_{n\mathbf{k}}|\partial_{\bfk} u_{n\mathbf{k}}},
\end{equation}
and the components of the Berry curvature tensor of band $n$ are given by:
\begin{equation}
\Omega^n_{ij}(\mathbf{k})=-2{\rm Im}\Braket{\partial_{k_i} u_{n\mathbf{k}}|
     \partial_{k_j} u_{n\mathbf{k}}}.
\end{equation}
The components of the Berry curvature itself are sometimes written as a vector
\begin{equation}
\Omega^n_{i}(\mathbf{k}):=\bfOmega^n_{i}(\mathbf{k})=(1/2)\epsilon_{lmi}\Omega^n_{lm}(\mathbf{k})=
-{\rm Im}\Braket{\partial_{\bfk} u_{n\mathbf{k}}|\times|
\partial_{\bfk} u_{n\mathbf{k}}}.
\end{equation}
The driving force behind the dynamics of an electron residing at a certain $\mathbf{k}$-point
and band $n$ could be an external electric or magnetic field as well as dependence of the Hamiltonian on another parameter, which, in a localized picture, 
cause the motion of an electron along certain orbits in $\mathbf{k}$-space and $\bfr$-space, as we shall see in detail in the following. The perturbation theory expression 
for the $\bfk$-space Berry curvature, looking at Eq.~(\ref{l-curv}), can be written as:
\begin{equation}
\Omega^n_{ij}(\bfk)=-{\rm 2\,Im}\sum_{m\neq n}\frac{\Braket{u_{n\bfk}|\partial_{k_i}H_{\bfk}|
u_{m\bfk}}\Braket{u_{m\bfk}|\partial_{k_j}H_{\bfk}|
u_{n\bfk}}}{(\varepsilon_{n\bfk}-\varepsilon_{m\bfk})^2}.
\end{equation} 

We will also consider a situation, in which, besides the dependence on $\bfk$, the Hamiltonian
of the system depends at the same time on another (multi-dimensional) parameter $\lambda$, that is, $H=H(\bfk,\lambda)$. Generally speaking, 
the Berry curvature form in this extended $(\lambda,\bfk)$ space has components, which we call
$\Omega^n_{\bfk\bfk}\equiv\Omega^n_{\bfk}$ and $\Omega^n_{\lambda\lambda}\equiv\Omega^n_{\lambda}$ and which
are expressed in terms of derivatives
of $\unkl$ with respect to only $\bfk$ or $\lambda$, respectively, according to (\ref{Berry-C}). 
However, there is also the component
of the Berry curvature form, which involves both $\lambda$- and $\bfk$-derivatives:
\begin{equation}
\Omega^n_{\lambda\bfk}=-2{\rm Im}\Braket{\partial_{\bfk}\unkl|\partial_{\lambda}\unkl}=
-{\rm 2\,Im}\sum_{m\neq n}\frac{\Braket{u_{n\bfk}|\partial_{\bfk}H_{\bfk}|
u_{m\bfk}}\Braket{u_{m\bfk}|\partial_{\lambda}H_{\bfk}|
u_{n\bfk}}}{(\varepsilon_{n\bfk}-\varepsilon_{m\bfk})^2}.
\end{equation}
We call this part of the Berry curvature  the {\it mixed Berry curvature}, and we will discuss it in detail in different contexts in the following sections. 

\subsection{Electric polarization}

The example of the electric polarization, besides being of great importance in solid state physics, will
allow us to delve into the concept of extended parameter space of reciprocal $\bfk$-vectors
as well as external parameter $\lambda$, with the generic
$(\bfk,\lambda)$-dependence of the Hamitonian and extension of the Berry curvature to higher dimensions.\iffindex{electric polarization}

Within the Born-Oppenheimer approximation, which assumes that the motion of electrons
is much faster than the slow motion of the ions, we can separate the ionic and electronic
terms in the charge density. The ionic charge density is simply given by sum of point
charges at the positions of the atoms, while we concentrate further on the electronic contribution to the polarization.  Within the
single-particle picture the electronic part of the charge
density is given by:
\begin{equation}
\rho^{\lambda}(\mathbf{r})= \sum_{n\le M}|\psi_n^{\lambda}(\mathbf{r})|^2,
\end{equation}
where $\lambda$ is an external parameter which for example specifies the atomic displacements,
$M$ is the highest occupied level and $\psi_n^{\lambda}$ are the single-particle states 
of the {\it finite} crystal:
\begin{equation}\label{l-potential}
\left(-\frac{\nabla^2}{2}+V^{\lambda}(\mathbf{r})\right)\psi_n^{\lambda}(\mathbf{r})=
\varepsilon_n^{\lambda}\psi_n^{\lambda}(\mathbf{r}).
\end{equation} 
For a finite sample with the volume $V_c$ the electronic part of the
electric polarization is given by
\begin{equation}\label{p-finite-sample}
\mathbf{P}^{\lambda}
= \frac{1}{V_c}\int_{\mathbb{R}_3} 
\mathbf{r}\rho^{\lambda}(\mathbf{r})\,d\mathbf{r}=
 \frac{1}{V_c}\int_{\rm sample} 
\mathbf{r}\rho^{\lambda}(\mathbf{r})\,d\mathbf{r},
\end{equation}
while its derivative  with 
respect to $\lambda$ expressed in terms of the matrix elements of the 
$\mathbf{r}$-operator is well-defined: 
\begin{equation}
\pollader=\frac{1}{V_c}\sum_{n\le M}\left(\Braket{\partial_{\lambda}\psi_n^{\lambda}|\mathbf{r}|\psi_n^{\lambda}}+\Braket{\psi_n^{\lambda}|\mathbf{r}|
\partial_{\lambda}\psi_n^{\lambda}}\right)=
\frac{1}{V_c}\,{\rm 2\,Re}\sum_{n\le M}\Braket{\partial_{\lambda}\psi_n^{\lambda}|\mathbf{r}|\psi_n^{\lambda}}.
\end{equation}
Consider now such a small change in $\lambda$ so that the change in the potential can be 
treated within the perturbation theory. In this case the derivative of the wave function
$\partial_{\lambda}\psi_n^{\lambda}$ in terms of other states within the first order
perturbation theory is given by Eq.~(\ref{pert-theo}) (we will soon
see that $\pollader$ is a gauge-invariant quantity and thus we can safely choose the parallel
transport gauge), which leads to the following result:
\begin{equation}
\pollader=\frac{1}{V_c}\,{\rm 2\,Re}\sum_{n\le M}\sum^{\infty}_{m\ne n}
          \frac{\Braket{\psi_n^{\lambda}|\mathbf{r}|
\psi_m^{\lambda}}\Braket{\psi_m^{\lambda}|\partial_{\lambda}H^{\lambda}|\psi_n^{\lambda}}}{\varepsilon_n^{\lambda}-\varepsilon_m^{\lambda}}.
\end{equation}
In case of a finite crystal the regions where the wavefunctions and the potential
are non-zero can be considered finite, thus the matrix elements of the position operator
in the expression above are well-defined. We will rewrite them now, however, using the
identity valid for $m\ne n$ (again properly defined only for finite samples)
\begin{equation}
\Braket{\psi_n^{\lambda}|\mathbf{r}|\psi_m^{\lambda}}=
-i\frac{\Braket{\psi_n^{\lambda}|\mathbf{p}|\psi_m^{\lambda}}}
{\varepsilon_n^{\lambda}-\varepsilon_m^{\lambda}},
\end{equation}
arriving at an alternative expression for the derivative of the polarization in a 
finite crystal:
\begin{equation}\label{finite}
\pollader=\frac{1}{V_c}\,{\rm 2\,Im}\sum_{n\le M}\sum^{\infty}_{m\ne n}
\frac{\Braket{\psi_n^{\lambda}|\mathbf{p}|
\psi_m^{\lambda}}\Braket{\psi_m^{\lambda}|\partial_{\lambda}
H^{\lambda}|\psi_n^{\lambda}}}
{(\varepsilon_n^{\lambda}-\varepsilon_m^{\lambda})^2}.
\end{equation}

This expression is  also well-defined in an infinite periodic crystal in terms of the Bloch orbitals
meaning that the derivative of the polarization represents a pure bulk property, free from the 
dependence on the termination of the crystal and physics at the surfaces. 
The expression for the $\pollader$ given as a sum of the matrix elements over occupied and 
unoccupied states integrated over the Brillouin zone constitutes a key result of the modern 
theory of electric polarization~\cite{resta1992}:
\begin{equation}\label{kubo}
\pollader=\frac{1}{(2\pi)^3}\sum_{n\le M}\sum^{\infty}_{m>M}{\rm 2\,Im}\intbz
\frac{\langle\pnkl|\mathbf{p}|\pmkl\rangle\langle\pmkl|\partial_{\lambda}
H^{\lambda} |\pnkl\rangle}{(\enla-\emla)^2}.
\end{equation}
Using the latter expression, 
one can show the Berry phase nature of the electric polarization.
Let us rewrite Eq.~(\ref{kubo}) in terms of the periodic functions $\unkl$, rather that
$\pnkl$, by noticing that: 
\begin{eqnarray}\label{Eq2}
\Braket{\psi_{n\mathbf{k}}^{\lambda}|\mathbf{p}|\psi_{m\mathbf{k}}^{\lambda}}=
\Braket{\unkl|\left[\partial_{\bfk},H^{\lambda}_{\bfk}\right]|\umkl}=
\Braket{\unkl|\partial_{\bfk} H^{\lambda}_{\bfk}|\umkl},\\
\Braket{\psi_{n\mathbf{k}}^{\lambda}|\partial_{\lambda} H^{\lambda}|
\psi_{m\mathbf{k}}^{\lambda}}=\Braket{\unkl|\left[\partial_{\lambda},
H^{\lambda}(\mathbf{k})\right]|\umkl}=\Braket{\unkl|\partial_{\lambda} H^{\lambda}_{\bfk}
|\umkl}.
\end{eqnarray}
In turn,
\begin{eqnarray}\label{Eq3}
\Bra{u_{n\mathbf{k}}^{\lambda}}\partial_{\mathbf{k}}H^{\lambda}_{\bfk}\Ket{u_{m\mathbf{k}}^{\lambda}} = 
(\varepsilon^{\lambda}_{\nkay}-\varepsilon_{\mkay}^{\lambda})
\Braket{\partial_{\mathbf{k}}u_{n\mathbf{k}}^{\lambda}|u_{m\mathbf{k}}^{\lambda}},\\
\Bra{u_{n\mathbf{k}}^{\lambda}}\partial_{\lambda}H^{\lambda}_{\bfk}\Ket{u_{m\mathbf{k}}^{\lambda}} = 
(\varepsilon^{\lambda}_{\nkay}-\varepsilon_{\mkay}^{\lambda})
\Braket{\partial_{\lambda}u_{n\mathbf{k}}^{\lambda}|u_{m\mathbf{k}}^{\lambda}}.
\end{eqnarray}
We can use these relations to reduce the band summation
in Eq.~(\ref{kubo}) to occupied states only~\cite{King-Smith}:
\begin{equation}\label{kubo-occ}
\pollader=\frac{1}{(2\pi)^3}\,{\rm 2\,Im}\sum_{n\le M}\intbz
\Braket{\partial_{\bfk}\unkl|\partial_{\lambda}\unkl} = 
- \frac{1}{(2\pi)^3}\sum_{n\le M}\intbz \, \Omega^n_{\lambda\bfk}.
\end{equation}
Thus, the derivative of the polarization with respect to $\lambda$, Eq.~(\ref{kubo-occ}), is 
the Brillouin zone integral of the mixed Berry curvature with respect to $\bfk$ and $\lambda$
in the $(\lambda,\bfk)$-space summed over all occupied bands. 
Since for a non-degenerate band all components of the curvature form
are gauge invariant, we conclude that $\pollader$ is a gauge invariant quantity.

We can now consider the change in $\polla$ as $\lambda$ is varied between, say,
0 and 1, under the condition that at each point along this interval our system stays
insulating. This change of the polarization during the $\lambda$-evolution of the system
can be expressed as: 
\begin{equation}
\depol=\int_{0}^{1}\pollader d\lambda,
\end{equation}
where we can substitute now equation~(\ref{kubo-occ}) and get:
\begin{equation}\label{delta-occ}
\depol=\frac{1}{(2\pi)^3}\,{\rm 2 Im}\sum_{n\le M}\int_{0}^{1}\intbz d\lambda
\Braket{\partial_{\bfk}\unkl|\partial_{\lambda}\unkl}
\end{equation}
In the special case of a one-dimensional lattice with a lattice constant $a$ and
BZ$=[-\frac{\pi}{a},\frac{\pi}{a}]$ we rewrite the change in polarization as:
\begin{equation}\label{1D}
\Delta P=\frac{1}{\pi}\,{\rm Im}\sum_{n\le M}\int_{-\frac{\pi}{a}}^
{\frac{\pi}{a}}\int_{0}^{1} dk\,d\lambda \Braket{\partial_{k}
u_{nk}^{\lambda}|\partial_{\lambda} u_{nk}^{\lambda}}.
\end{equation}

\begin{figure}[t!]
\begin{center}
\includegraphics[width=0.95\textwidth]{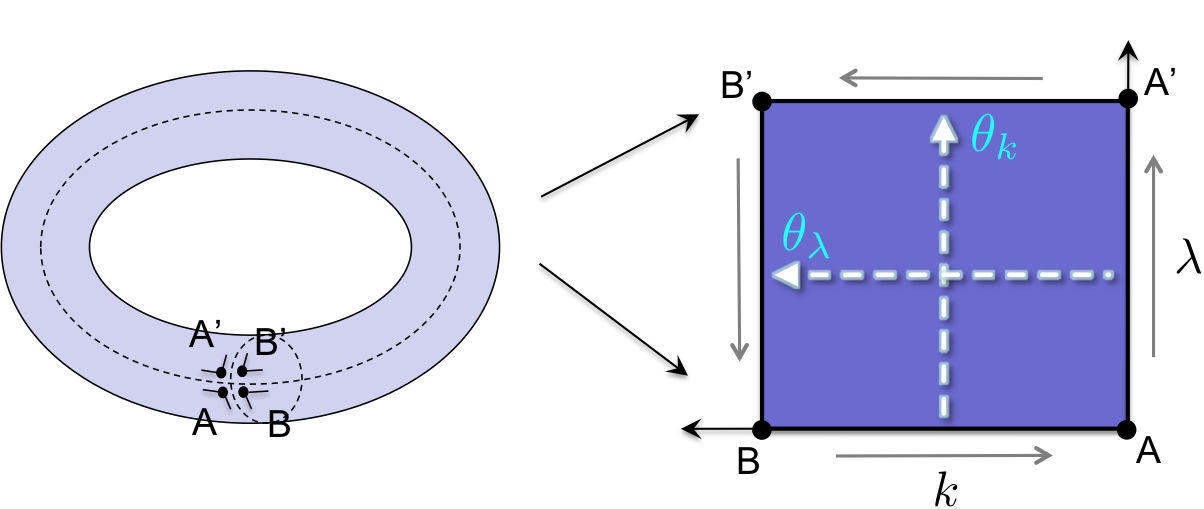}
\end{center}
\caption{\label{pic_intro} Left: torus of states $u_{n\bfk}^{\lambda}$ where $\lambda$ runs from 0 to 1
with $H^{\lambda=0}_\bfk=H^{\lambda=1}_\bfk$, and $\bfk$ runs in the first Brillouin zone. 
Right: cutting the torus on the left into a square avoiding going through the equators of the torus. This allows
to choose smooth and unique gauge everywhere in the square and at its boundaries. For details see text.}
\end{figure}

{\it \textbf{Topological meaning of electric polarization.}}\\
While the geometrical meaning of $\Delta P$ as a property related to the Berry curvature in $\bfk$-space
has been clarified, a way of looking at the $\Delta P$ as a {\it topological} property of the occupied states requires
to make a compact topological manifold without boundary out of the Brillouin zone. An intuitive way of doing so would be
to glue the edges of the BZ which differ by a reciprocal $\bfG$ vector, together,~i.e., to identify $u_{n\bfk}$
with $u_{n\bfk+\bfG}$. This however cannot be done by just putting $u_{n\bfk}=u_{n\bfk+\bfG}$, since 
the Hamiltonians at $\bfk$ and $\bfk+\bfG$ are not the same, but are related as follows: 
\begin{equation}
H_{\bfk+\bfG}=e^{-i\bfG\bfr}H_{\bfk}\,e^{i\bfG\bfr}. 
\end{equation}
From the latter relation we can conclude that the set of instantaneous solutions of $H_{\bfk+\bfG}$ can be always chosen to be as follows:
\begin{equation}
\{ u_{n\bfk+\bfG}\} = e^{-i\bfG\bfr}\{e^{i\theta_n}u_{n\bfk}\},
\end{equation}
where $\theta_n$ are band-dependent constants which stand for the obvious $\mathbb{U}(1)$ gauge freedom
which we considered previously. It can be shown straightforwardly that the Berry phase $\gamma_n$ computed
along a path which connects the $\bfk$ and $\bfk+\bfG$ points is a gauge-invariant quantity, without the path
of integration being formally closed, and the whole geometrical machinery we presented in section 2 can be also 
applied here. Indeed, imagine two families of Hamiltonians, $H(\lambda)$ and $H'(\lambda)$,  which differ by a
{\it constant} $\lambda$-independent unitary transformation:
\begin{equation}
H'(\lambda) = UH(\lambda)U^{\dagger},
\end{equation}
with $U$ and $H(\lambda)$ not necessarily commuting. Then it is clear that if at a point $\lambda$ the set of 
instantaneous solutions of $H(\lambda)$ is $\{\Ket{n\lambda}\}$, then the set $\{  \Ket{n\lambda}' \}=\{  U\Ket{n\lambda} \}$
presents a set of instantaneous solutions of $H'(\lambda)$. If in the vicinity of certain $\lambda$ function $\Ket{n\lambda}$ is smooth, so will be $\Ket{n\lambda}'$. In terms of a gauge transformation, the two sets
are connected by a trivial, $\lambda$-independent gauge transformation. The connections of $H$ and $H'$ can be also
compared:
\begin{equation}\label{eightythree}
\calA'^n = i\braket{n\lambda'|\partial_{\lambda}|n\lambda'}d\lambda= i\braket{Un\lambda|\partial_{\lambda}|Un\lambda}d\lambda=
i\braket{n\lambda|\partial_{\lambda}|n\lambda}d\lambda = \calA^n.
\end{equation}
This means that the Berry phase of any closed path $\calC$ for both Hamiltonians is the same. This is actually true even for 
any path $\calC$ which is not closed. In case of Bloch electrons, the role of $U$ is played by $e^{-i\bfG\bfr}$.

The most convenient way to deal with the topological properties of the BZ  is to consider
what we call the {\it folded BZ}. At each point $\bfk$ in the first BZ we consider the {\it equivalence classes} of lattice periodic 
functions which differ from each other by $e^{i\bfG\bfr}$, where $\bfG$ is a reciprocal lattice vector. The equivalence
class which corresponds to a certain $u_{n\bfk}$ is given by:
\begin{equation}
\hat{u}_{n\bfk}=\{   e^{-i\bfG\bfr} u_{n\bfk},\,\forall {\bfG} \}.
\end{equation}
Physically, $\hat{u}_{n\bfk}$ is an object which accumulates all instantaneous solutions at points $\bfk+\bfG$ in the reciprocal space and 
folds them into the first BZ. $u_{n\bfk}$ is called the {\it representative} of this class. Respectively, we can consider the equivalence classes of the Hamiltonians,
\begin{equation}
\hat{H}_{\bfk} = \{ e^{-i\bfG\bfr} H_\bfk \, e^{i\bfG\bfr},\,\forall {\bfG}\},
\end{equation}
for which corresponding $\hat{u}_{n\bfk}$ are the instantaneous solutions:
\begin{equation}
\hat{H}_\bfk \hat{u}_{n\bfk} = \varepsilon_{n\bfk}\hat{u}_{n\bfk}.
\end{equation}
Importantly, as opposed to $H_{\bfk}$, the corresponding equivalence class $\hat{H}_\bfk$ is periodic in reciprocal space, since
$\hat{H}_{\bfk+\bfG}=\hat{H}_\bfk$. 
It is easy to show that the Berry phase
theory from above can be transparently formulated in terms of $\hat{u}_{n\bfk}$'s rather then $u_{n\bfk}$'s. If the gauge
freedom in the choice of $\hat{u}_{n\bfk}$'s is introduced as
\begin{equation}
\hat{u}_{n\bfk}' = e^{i\theta_n(\bfk)}\hat{u}_{n\bfk} := \{ e^{i\theta_n(\bfk)}  e^{-i\bfG\bfr} u_{n\bfk},\,\forall {\bfG} \},
\end{equation} 
then we can show, using (\ref{eightythree}), that the Berry connection $\hat{\calA}^n$ can be defined uniquely for $\hat{u}_{n\bfk}$ as
\begin{equation}
\hat{\calA}^n=i\braket{\hat{u}_{n\bfk}|\partial_{\bfk}|\hat{u}_{n\bfk}}d\bfk:= i\braket{u_{n\bfk}|\partial_{\bfk}|u_{n\bfk}}d\bfk,
\end{equation}
with gauge transformation reading in analogy to (\ref{gauge-A}):
\begin{equation}
\hat{\calA}'^n = \hat{\calA}^n - d\theta_n.
\end{equation}
Finally, in our construction of the folded BZ we identify the BZ zone boundaries which differ from each other by any $\bfG$, 
which can be done since $\hat{H}_\bfk$ is a periodic function in reciprocal space.
In case of a 2D BZ this looks like  a torus, see Fig.~\ref{pic_intro}, and results in a topological construction of a closed compact manifold. As we remember from the example of spin-$\frac{1}{2}$ in magnetic field
the Chern number was ultimately related to our inability of constructing a global smooth and unique gauge on $S^2$. The same is true for our folded BZ, as we shall see below. If, starting from a point $\bfk$ in a folded BZ, we could go around our
torus along one of the diameters smoothly and uniquely on the whole closed path, obviously, we would arrive at the condition
that:
\begin{equation}
\hat{u}_{n\bfk}\longrightarrow\hat{u}_{n\bfk}\quad \Leftrightarrow \quad u_{n\bfk+\bfG}=e^{-i\bfG\bfr}u_{n\bfk}, \, \forall \bfG.
\end{equation}
The choice of such a gauge in the whole folded BZ, if it is possible to make, is called the {\it periodic gauge}. If the Chern number of
our manifold is non-zero, such a choice is impossible to make, and, generally speaking:
\begin{equation}\label{EQUIV}
\hat{u}_{n\bfk}\longrightarrow e^{i\theta_n}\hat{u}_{n\bfk},
\end{equation}
where (generally $\bfk$-dependent) $\theta_n$ is not necessarily a multiple of $2\pi$. Next, lets analyze in detail the
relation between the Chern number and the phase $\theta_n$ for the case of electric polarization (\ref{1D}). \\

{\it \textbf{Chern number and the 2-point formula.}}\\
Let us consider the case of $H(\lambda=0) = H(\lambda=1)$. For the $(\lambda,k)$ situation of Eq.~(\ref{1D}) the general relation between the instantaneous solutions at the opposite
sides of the square read, analogously to (\ref{EQUIV}):
\begin{eqnarray}\label{boundary-cond}
u_{k}^{\lambda=1} &=& e^{i\theta_{k}}u_{k}^{\lambda=0},\\\label{boundary-cond2}
u_{\frac{\pi}{a}}^{\lambda} &=& e^{-iGx}\,e^{i\theta_{\lambda}}u_{-\frac{\pi}{a}}^{\lambda},
\end{eqnarray}
where we skip for simplicity the band index $n$. 
Let us now try to evaluate the $\Delta P$. Following Fig.~\ref{pic_intro} we can cut the torus such that on the boundaries of
the  square $AA'B'B$ we have a smooth choice of the $u_{k}^{\lambda}$. This is equivalent to the reasonable assumption that all the ``non-smoothness" has been restricted to the infinitesimally small ``belts" around the two equators of the torus, which  
are analogous to an infinitesimally small ``belt" around the equator of the $S^2$ in case of spin-$\frac{1}{2}$ in magnetic field. Then the integral of the mixed Berry curvature over the torus, given by Eq.~(\ref{1D}), can be rewritten as an integral of the Berry connection along the path $\mathcal{C}=AA'B'B$, Fig.~\ref{pic_intro}:
\begin{equation}\label{square-A}
\Delta P = \frac{1}{2\pi}\left(  \int_{0}^1d\lambda\left[ \calA_{\lambda}(\lambda,-\frac{\pi}{a}) - \calA_{\lambda}(\lambda,\frac{\pi}{a}) \right]     +   \int_{-\pi/a}^{\pi/a}dk\left[ \calA_{k}(1,k) - \calA_{k}(0,k) \right] \right),
\end{equation}
where $\calA_{\lambda}=i\Braket{\unkl|\partial_{\lambda}\unkl}$ and 
$\calA_k=i\Braket{\unkl|\partial_{\bfk}\unkl}$ are the components of the Berry connection. Using Eqs.~(\ref{boundary-cond})
and (\ref{boundary-cond2}), it is easy to see that the expression for $\Delta P$ is reduced to:
\begin{equation}
\Delta P = \frac{1}{2\pi}\left( \int_0^1 \frac{\partial\theta_{\lambda}}{\partial\lambda}d\lambda -
\int_{-\pi/a}^{\pi/a} \frac{\partial\theta_{k}}{\partial k}dk \right).
\end{equation}
Since the functions $\theta$ are smooth along the boundary, we come to the conclusion that
\begin{equation}
\Delta P = \frac{\theta_{\lambda}(1) - \theta_{\lambda}(0) - \theta_k(\pi/a) + \theta_k(-\pi/a)}{2\pi}.
\end{equation}
On the other hand, according to Eqs.~(\ref{boundary-cond})
and (\ref{boundary-cond2}), upon the $A\rightarrow A' \rightarrow B' \rightarrow B$ evolution the wavefunction 
$u_{-\pi/a}^{\lambda=0}$ is returned into the wavefunction $e^{i(\theta_{\lambda}(1) - \theta_{\lambda}(0) - \theta_k(\pi/a) + \theta_k(-\pi/a))}u_{-\pi/a}^{\lambda=0}$. Since we chose our gauge smoothly along this path, 
\begin{equation}
u_{-\pi/a}^{\lambda=0} = e^{i(\theta_{\lambda}(1) - \theta_{\lambda}(0) - \theta_k(\pi/a) + \theta_k(-\pi/a))}u_{-\pi/a}^{\lambda=0},
\end{equation}
meaning that $\theta_{\lambda}(1) - \theta_{\lambda}(0) - \theta_k(\pi/a) + \theta_k(-\pi/a)$ has
to be a multiple of $2\pi$, and $\Delta P$ has to be an integer. In fact, it is the first Chern number of
the system, as we have seen in the previous section:
\begin{equation}
\Delta P = C \quad \longrightarrow \quad \Delta P = \sum_n C_n,
\end{equation}
where $C_n$ is the first Chern number of band $n$ in the $(\lambda,k)$ space and the arguments above
present the proof of the quantization of the Chern number for this particular situation. This proof  is however 
more general than the one we provided for the spin-$\frac{1}{2}$ in
magnetic field, since it does not rely on the precise expressions for the Berry connection~(\ref{Sp-A}). Depending on the
Hamiltonian of the system, $H_{k}^{\lambda}$, the Chern number of the system, which {\it uniquely} in
mathematical sense depends on the system, can be either zero, or it can be non-zero. In the latter case
we say that we encounter a situation of a {\it Chern insulator}, as opposed to the trivial insulator with
the Chern number zero. In the previous section we provided an example of a Hamiltonian which is a
Chern insulator, namely, a spin-$\frac{1}{2}$ particle in magnetic field, in which case the sphere $S^2$ played
the role of the $(\lambda,k)$ space. In case when the system is not a Chern insulator the smooth and unique choice of the wavefunction
can be found everywhere on the torus, which corresponds to the case of  $\theta_k\equiv\theta_{\lambda}\equiv 0$ and
$u_{k}^{\lambda=1} =u_{k}^{\lambda=0}, u_{\pi/a}^{\lambda} = e^{-iGx}u_{-\pi/a}^{\lambda}$, as discussed above.
In the next section we elaborate in more detail on Chern insulators
in two-dimensional reciprocal space.

Let us assume that we can choose the periodic gauge in the reciprocal space for each $\lambda$. 
 In this case $\theta_{\lambda}\equiv 0$ and the
 change of the polarization is given by:
\begin{equation}\label{X}
\Delta P = \frac{\theta_k(-\pi/a) - \theta_k(\pi/a)}{2\pi}=-\int_{-\pi/a}^{\pi/a} \frac{\partial\theta_{k}}{\partial k}dk.
\end{equation}
We can even go further back to write the so-called {\it 2-point formula} for the change of polarization:
\begin{equation}\label{2point}
\Delta P = P_1 - P_0 = \frac{i}{2\pi}\int_{-\pi/a}^{\pi/a}dk\,\Braket{u^{\lambda}_{k}|\partial_{k}u^{\lambda}_{k}}^{\lambda=1}
-\frac{i}{2\pi}\int_{-\pi/a}^{\pi/a}dk\,\Braket{u^{\lambda}_{k}|\partial_{k}u^{\lambda}_{k}}^{\lambda=0}.
\end{equation}
The latter equation absolutely correctly gives the value of $\Delta P$, if we remember that we started
from Eq.~(\ref{X}). What is meant is that very often, especially in practical calculations of the polarization, 
the expresion (\ref{2point}) is considered separately without its reference to Eq.~(\ref{X}). This means that the sets
of $\{  u_{nk}^{\lambda=0} \}$ and $\{  u_{nk}^{\lambda=1} \}$ wavefunctions are calculated separately omitting the $\lambda$-evolution, and the value of $\Delta P $ is then calculated as in (\ref{2point}). In this way we do not
built explicitly a unique smooth function $\theta_k$ and it can in principal take any values. Applying 
expression (\ref{2point}) properly would mean, given a certain set of $u_{nk}$ in the BZ for $\lambda=0$,
to go smoothly by hand from $u_{nk}^{\lambda=0}$ across $u_{nk}^{\lambda}$ to construct $u_{nk}^{\lambda=1}$,
and to use such constructed  $u_{nk}^{\lambda=1}$ in (\ref{2point}).  
If we do not do this, then the only thing
we know is that due to periodic boundary conditions $e^{i\theta_k(\pi/a)}=e^{i\theta_k(-\pi/a)}$, meaning
that the difference of $\theta_k(-\pi/a) - \theta_k(\pi/a)$ is defined up to a multiple of $2\pi$, with $\Delta P$
as an integer, but undetermined in value. The quantity (general to multi-dimensional $\lambda$ and $\bfk$) 
\begin{equation}
\polla=\frac{i}{(2\pi)^3}\sum_{n\le M}\intbz
\Braket{\unkl|\partial_{\bfk}\unkl}
\end{equation}
is called the {\it electric polarization} and the two-point formula for $\Delta \mathbf{P}$ can be generalized
in exact analogy to Eq. (\ref{2point}) to higher dimensions. It is straightforward to show that if $H(\lambda=0)\neq
H(\lambda=1)$, Eq. (\ref{square-A}) is still valid. This means, that although $\theta_k$ cannot be appropriately
defined anymore, given the possibility of a periodic gauge in $\bfk$-space for each $\lambda$, and a smooth
connection between  $u_{nk}^{\lambda=0}$ and $u_{nk}^{\lambda=1}$, the two-point formula can be also
applied, although the value of $\Delta P$ is not anymore quantized. \\

{\it \textbf{Adiabatic pumping and velocity of Bloch electrons}.}\\
Evaluation of the velocity of electronic states requires going beyond the adiabatic approximation for the evoluted
wavefunction (\ref{adiabatic1}-\ref{adiabatic2}). Up to first order in transition frequencies (\ref{T-coeff}) equations 
(\ref{QUASI}) can be solved to yield:
\begin{equation}\label{previous}
\Ket{\psi(t)}=e^{-i\int_0^{t}\varepsilon_n(\tau)d\tau}\left[     
\Ket{n\lambda(t)}-i\sum_{m\neq n}\frac{\braket{m\lambda(t)|\partial_t|n\lambda(t)}}{\varepsilon_n(t)
-\varepsilon_m(t)}\Ket{m\lambda(t)}
\right].
\end{equation}
The latter expression goes beyond the adiabatic approximation in that it also includes the 
``smearing" of the wavefunction which was initially in state $n$, over other
eigenstates of the Hamiltonian during the time-propagation. In deriving (\ref{previous}) the parallel transport gauge (\ref{PT-nl})
 was used, which locally eliminates the geometric phase contribution to the evolution of $\psi$. This is fine, however,
 since the final expression will be locally gauge-invariant. Let us now try to evaluate the velocity of
electronic states in a crystal during time-evolution given by change in parameter $\lambda$, which could be time itself. Namely, 
at time $t=0$ at a certain $\bfk$ we start with a wavefunction $\psi_{n\mathbf{k}}^{\lambda(t=0)}$ which is analogous to 
state $\Ket{n\lambda(0)}$ from above, and we evaluate the velocity of the evoluted wavefunction at infinitesimally
small time $t>0$. It is most convenient to assume at the moment that $\lambda$ enters the Hamiltonian according
to (\ref{l-potential}) via the crystal potential, in which case $\bfk$ stays constant as $\lambda$ is changed.
To distinguish the properly evoluted states from the instantaneous states $\psi_{n\mathbf{k}}^{\lambda(t)}$
and $u_{n\mathbf{k}}^{\lambda(t)}$ (playing the role of $\Ket{n\lambda(t)}$ above), we denote them by $\tilde{\psi}_{n\mathbf{k}}^{\lambda}$
and $\tilde{u}_{n\mathbf{k}}^{\lambda}$. Then the average velocity of state $\tilde{\psi}_{n\mathbf{k}}^{\lambda}$
at point $\lambda=\lambda(t)$ is given by (using also (\ref{Eq2})):
\begin{equation}
\bfv_{n\mathbf{k}}^{\lambda}=
\Braket{\tilde{\psi}_{n\mathbf{k}}^{\lambda}|i\left[ H^{\lambda}, \mathbf{r} \right]
|\tilde{\psi}_{n\mathbf{k}}^{\lambda}}=
\Braket{\tilde{u}_{n\mathbf{k}}^{\lambda}|\partial_{\mathbf{k}} H^{\lambda}_{\mathbf{k}}|\tilde{u}_{n\mathbf{k}}^{\lambda}}.
\end{equation}
On the other hand from (\ref{previous}) it follows that: 
\begin{equation}
\ket{\tilde{u}_{n\mathbf{k}}^{\lambda}}=e^{-i\int_0^{t}\varepsilon^{\lambda(\tau)}_{n\mathbf{k}}\,d\tau}\left[ 
\ket{u_{\nkay}^{\lambda}} - i \sum_{m\neq n}\ket{u_{\mkay}^{\lambda}}
\frac{\Braket{u_{\mkay}^{\lambda}|\partial_{\lambda}u_{\nkay}^{\lambda}}}{\varepsilon^{\lambda}_{\nkay}
-\varepsilon_{\mkay}^{\lambda}}\,\dot{\lambda}
\right],
\end{equation}
where $\dot{\lambda}=\partial_t\lambda(t)$. Then,
\begin{equation}
\bfv_{\nkay}^{\lambda} = \partial_{\bfk}\varepsilon_{\nkay}^{\lambda} + 
2{\rm Im} \sum_{m\neq n}\Bra{u_{n\mathbf{k}}^{\lambda}}\partial_{\mathbf{k}}H^{\lambda}_{\mathbf{k}}\Ket{u_{m\mathbf{k}}^{\lambda}}\frac{\Braket{u_{\mkay}^{\lambda}|\partial_{\lambda}u_{\nkay}^{\lambda}}}{\varepsilon^{\lambda}_{\nkay}-\varepsilon_{\mkay}^{\lambda}}\,\dot{\lambda}. 
\end{equation}
Using Eqs.~(\ref{Eq3}) from before, 
we arrive at the following two equivalent expressions for the velocity of the state
\begin{equation}\label{106}
 \bfv_{\nkay}^{\lambda}  =\partial_{\bfk}\varepsilon_{\nkay}^{\lambda} + 2{\rm Im}\sum_{m \neq n}\frac{\Braket{ u_{\nkay}^{\lambda}| \partial_{\bfk}H^{\lambda}_{\bfk}| u_{\mkay}^{\lambda}}\Braket{u_{\mkay}^{\lambda} |\partial_{\lambda}H^{\lambda}_{\bfk} | u_{\nkay}^{\lambda}}}{(\varepsilon^{\lambda}_{\nkay}-\varepsilon_{\mkay}^{\lambda})^2}\,\dot{\lambda},
\end{equation}
or, according to (\ref{l-curv}),
\begin{equation}
 \bfv_{\nkay}^{\lambda}  = 
 \partial_{\bfk}\varepsilon_{\nkay}^{\lambda} - \Omega^n_{\lambda\bfk}\,\dot{\lambda}.
\end{equation}
In this expression the first term of the right hand side is the {\it group velocity}, which is present also when $\lambda$ is constant and the stationary state $\psi^{\lambda}_{n\bfk}$ evolves in time according to (\ref{stationary}). The change
in $\lambda$ gives, on the other hand, rise to the so-called {\it anomalous velocity}, which is expressed in terms of the mixed Berry curvature $\Omega^n_{\lambda\bfk}$. Clearly, the anomalous velocity of a state depends on how fast the parameter $\lambda$
is changed in time.
We can now evaluate the current density due to all occupied states changing of
$\lambda$ in time induces. It is given by:
\begin{equation}
\mathbf{J}^{\lambda}=-\frac{1}{(2\pi)^d}\,\dot{\lambda}\sum_{n\le M}\int_{\rm BZ}\Omega^n_{\lambda\bfk}\,d\bfk,
\end{equation}
while the contribution due to the group velocity of the states clearly vanishes, since the band energies are periodic functions of $\bfk$. The polarization charge pumped during the evolution from $\lambda_1$ to $\lambda_2$ is given by the time integral of the current density from above from $t_1$ to $t_2$ with $\lambda(t_1)=\lambda_1$ and $\lambda(t_2)=\lambda_2$, which amounts to:
\begin{equation}
\Delta \mathbf{P} = \int_{t_1}^{t_2}\mathbf{J}^{\lambda(t)}\,dt= -\frac{1}{(2\pi)^d}\sum_{n\le M}\int_{\lambda_1}^{\lambda_2}\int_{\rm BZ}\Omega^n_{\lambda\bfk}\,d\bfk d\lambda,
\end{equation}
which gives us exactly the change in the polarization as given by Eq.~(\ref{delta-occ}). This expression presents a theoretical justification for the interpretation of electric polarization in terms of experimentally measured charge current. In the special case of one dimension, as we have
seen previously, the transported through the system charge during cyclic adiabatic evolution
is thus quantized, which leads to the phenomenon of {\it adiabatic pumping}.\iffindex{adiabatic pumping}
\\

{\it \textbf{Symmetry properties of the Berry curvature}.}\\
Based on general symmetry arguments, it is straightforward to show that the Berry curvature in
$\bfk$-space obeys the following symmetry properties: (i) in the presence of time-reversal symmetry,
$\Omega^n(-\bfk)=-\Omega^n(\bfk)$, while (ii) in the presence of the space inversion symmetry,
 $\Omega^n(-\bfk)=\Omega^n(\bfk)$. This means that when both space and time inversion symmetry
 are present in a solid, the Berry curvature at each $\bfk$ is identically zero. In case of materials,
 which display non-zero electric polarization, the Berry curvature is non-trivial owing to the breaking
 of inversion symmetry. On the other hand, in materials which exhibit spontaneous magnetization,
 such as ferromagnets and antiferromagnets, the non-trivial Berry curvature is due to breaking
 of time-inversion symmetry. We will focus on the latter case in the next sections.

\subsection{Chern insulators and (quantum) anomalous Hall effect}
In the remainder of these notes we will call the {\it Chern insulator} a two-dimensional (2D) insulating solid with Hamitonian 
$H(\bfk)$ whose first Chern number, determined as an integral of the $\bfk$-space Berry curvature over the
Brillouin zone, is an integer {\it non-zero} number. The proof of the fact that the Chern number is integer
for a two-dimensional insulator we have provided in the previous section, where one has to replace $k$
with $k_x$ and $\lambda$ with $k_y$. The condition of the periodicity of the Hamiltonian with respect to
$\lambda$ which we used to glue the square in Fig.~\ref{pic_intro} into a torus can be satisfied in reciprocal space
in $k_y$-direction following the folding procedure we decribed above, which allows us to look at the 2D Brillouin 
zone as a torus. The condition that the Chern
number is non-zero means that the topological properties of our system are non-trivial, and that the 
wavefunction $u_{n{\bfk}}$ acquires a ``twist'' as we cross at least one of the equators, which manifests
in the non-zeroness of $\theta_{k_x}$ or $\theta_{k_y}$. Chern insulators present an example of a system
for which the periodic gauge in both $k$-directions cannot be found.\iffindex{Chern insulator}

One of the most remarkable properties of Chern insulators is the quantization of their
transverse Hall conductance. From the general linear
response formalism as well as from the semiclassical electron dynamics which we consider 
later, it follows that Hall conductance of a 2D system is given by:
\begin{equation}\label{2D-Hall}
\sigma_{xy}=\frac{1}{2\pi}\int_{\rm BZ}d\bfk\,\Omega_{xy}(\bfk) = \frac{1}{2\pi}\int_{\rm BZ}d\bfk\,\Omega_k = \sum_n C_n,
\end{equation}
where $C_n$ is the first Chern number of the occupied band $n$.
From the point of view of adiabatic pumping we considered previously, the quantization of the
Hall conductance is due to the quantized charge which is pumped through the one-dimensional
system along $x$ described by Hamiltonian $H(k_x,k_y)$ as the parameter $k_y$ is varied from one side
of the BZ to the other. This corresponds to the situation of varying $\lambda$ between 0 and 1 in
Eq. (\ref{1D}).

Let us take a look at the mechanisms which can lead to appearance of Chern insulators. For this 
purpose we consider first the case of only two bands in $\bfk$-space (keeping in mind for future
reference that the role of $\bfk$ can be replaced by any parameter). Neglecting the constant term,
which does not change the topological properties and just shifts the energy, a generic 2D Hamiltonian
reads:
\begin{equation}\label{111}
H(\bfk) = \bfd(\bfk)\cdot\bfsigma,
\end{equation}
where $\bfsigma$ is the vector of Pauli matrices. At each point in $\bfk$ the energetic spectum is
given by two eigenvalues $\varepsilon_{+}$ and $\varepsilon_{-}$ which are given by 
$\pm|\bfd(\bfk)|=\pm d(\bfk)$. Since the Berry curvature summed over both bands
 is always zero, we will consider only lowest of
the bands, $\varepsilon_{-}$. We also suppose that both bands are separated from each other by a
gap. By comparing  the Hamiltonian (\ref{111}) to that of spin-$\frac{1}{2}$ in magnetic field, 
it is clear that the physics of the problem is governed by the Dirac monopole. Indeed, the Berry phase which is accumulated when going along a closed loop $\mathcal{C}$ in $\bfk$-space can be easily related to that picked up
when going along a loop $\chi(\mathcal{C})$ on a unit sphere $S^2$, where $\chi$ maps $\bfk$
to the point $\bfd/d$ on $S^2$, see Fig.~\ref{Sphere}:
\begin{equation}\label{chi}
\chi: \bfk \longrightarrow \bfn=\bfd/d,\quad\bfn\in S^2.
\end{equation} 
The Berry curvature of the problem on $S^2$ is given by a field of the
Dirac monopole with quantized charge at the origin, where the two bands touch each other and $\varepsilon_{+}=\varepsilon_{-}$. As we remember from the considerations of $S^2$, for a given spin we cannot choose the smooth gauge of our wavefunctions in the BZ and we have to ``glue" them together at the diameter of the sphere. The Berry curvature in the $\bfk$-space is readily
obtained from the Berry curvature of the Dirac monopole multiplied with the Jacobian of $\chi$:
\begin{equation}\label{113}
\Omega_{xy}(\bfk)=-\,\bfn\cdot\left(   \partial_{k_x}\bfn \times \partial_{k_y}\bfn \right)/2 =
\bfOmega^{\sigma}_{\bfn}\cdot\left( \partial_{k_x}\bfd \times \partial_{k_y}\bfd   \right),
\end{equation}
where $\bfOmega^{\sigma}_{\bfn}$ is that given by Eq.~(\ref{Sp-A}) for $\sigma=-1$. As we learned from mathematics, the first Chern number obtained as an integral of the Berry curvature over a {\it compact} manifold such as a 2D BZ,
has to be an integer. For (\ref{113}) it is called the {\it winding number} and it stands for the number of times that the 
field $\bfd(\bfk)$ winds around the $S^2$ as $\bfk$ is varied, as can be intuitively understood and computed explicitly. \iffindex{winding number}

A good example when the integer quantization is violated is the massive Dirac Hamiltonian, for
which $d_x=k_x, d_y=k_y$ and $d_z=m$: 
\begin{equation}\label{114}
H(\bfk)=k_x\sigma_x+k_y\sigma_y+m\sigma_z.
\end{equation}
This Hamiltonian both with zero and non-zero $m$ is of great importance for studying topological
properties of solids. The Berry curvature can be evaluated analytically: 
\begin{equation}
\Omega_{xy}(\bfk)=m/
[ 2(m^2 + k^2)^{3/2}], 
\end{equation}
which leads to the following Hall conductance when integrated over the
whole infinite BZ: $\sigma_{xy}={\rm sgn}(m)/2$, leading thus to half-integer-quantized values.
 This seems to be in contradiction with our expectation of  integer quantization. The 
reason for this is that we have here a situation of a non-compact manifold for $\bfk$-space, namely, 
$\mathbb{R}^2$. By looking at the distribution of $\bfd$ over $\mathbb{R}^2$ we realize
that for a given $m$ the sign of $d_z$ remains constant and the vector $\bfd$ spans only half
of the $S^2$, leading to the so-called meron situation. This results in half-integer
conductance. In a realistic situation of a lattice which provides a periodic lattice potential and finite
band width, the BZ is compact, and the sign of $d_z$ changes at other points of the BZ, where
the bands bend down in order to assure a finite band width. Such regions provide the other half
of the conductance, vector $\bfd$ spans the other half of the sphere and the resulting 
conductance indeed becomes integer. The moral of this story is that in order to determine the
Chern number of a material, it is not enough to concentrate only on the Dirac-like localized points 
in the BZ which can provide very large contribution to the Berry curvature, and the whole band
structure of the crystal has to be considered. 

At the point $m=0$ in the Hamiltonian (\ref{114}) there is a point of degeneracy in the spectrum at $k_x=k_y=0$ and
the transition between two Chern insulator phases for $m<0$ and $m>0$ occurs.
Such situation is rather typical and can be generalized to the case of a 2D+1 Hamiltonian
$H(k_x,k_y,\lambda)=H(\bfkappa)$, where $\bfkappa=(k_x,k_y,\lambda)$ and $\lambda$ is
a certain parameter, such as $m$ in the Dirac model, values of hoppings in the lattice
Hamiltonian, value of the spin-orbit strength, exchange field, etc.  According to the theorem by Bellisard~\cite{Bellisard},
the change of the $(k_x,k_y)$ Chern number when going through a point of
degeneracy at a certain value of $\bfkappa^*$ (such as $(0,0,0)$ in the Dirac model) is given by the
so-called {\it Berry index}:
\begin{equation}
{\rm Indx_B} = \frac{1}{2\pi}\int_{S^2}\Omega_k(\bfkappa)d\bfkappa,
\end{equation}
where $S^2$ is an infinitesimally small sphere which encloses $\bfkappa^*$. Interestingly,
the Berry index determines the change in the 2D Hall conductance irrespective of the compactness
of the $\bfk$-space. E.g. for the massive Dirac Hamitonian (\ref{114}) the Berry index can be evaluated
to be $+1$ at the point $(k_x=0,k_y=0,m=0)$. Moreover, it can be shown that the value of
the Berry index is pre-determined by the band dispersion in the vicinity of $\bfkappa^*$~\cite{Chang}.
 Infinitesimally closely to $\bfkappa^*$ we can approximate our two-band Hamiltonian as (omitting
  again the constant energy term):
\begin{equation}
H(\bfkappa)  = h(\bfkappa-\bfkappa^*)\cdot\bfsigma.
\end{equation}
If the dispersion of $h(\bfkappa-\bfkappa^*)$ is linear, then the Berry index assumes the values
of $\pm 1$, while if it is quadratic, ${\rm Indx_B}$ is $\pm 2$. For the massive Dirac model (\ref{114})
Hamiltonian $h(\bfkappa-\bfkappa^*)$ is linear in $\bfkappa-\bfkappa^*$ which results in the change
of Hall conductance by $+1$ as the mass $m$ changes sign.
It is probably worthy to note here that when the role of the parameter $\lambda$ is played 
by the $k_z$ Bloch vector of a 3D Hamiltonian $H(k_x,k_y,k_z)$, such points of degeneracy
$\bfkappa^*=\bfk^*$
will be called here {\it Weyl points}. If such points happen to be present at the Fermi energy of a material
with no other bands crossing it, such material is called a {\it topological metal}, or {\it Weyl
semimetal}.\iffindex{Weyl semimetal} The physics of topological metals is an exciting emerging field of topological 
solid state physics. 

\begin{figure}[t!]
\begin{center}
\includegraphics[width=0.85\textwidth]{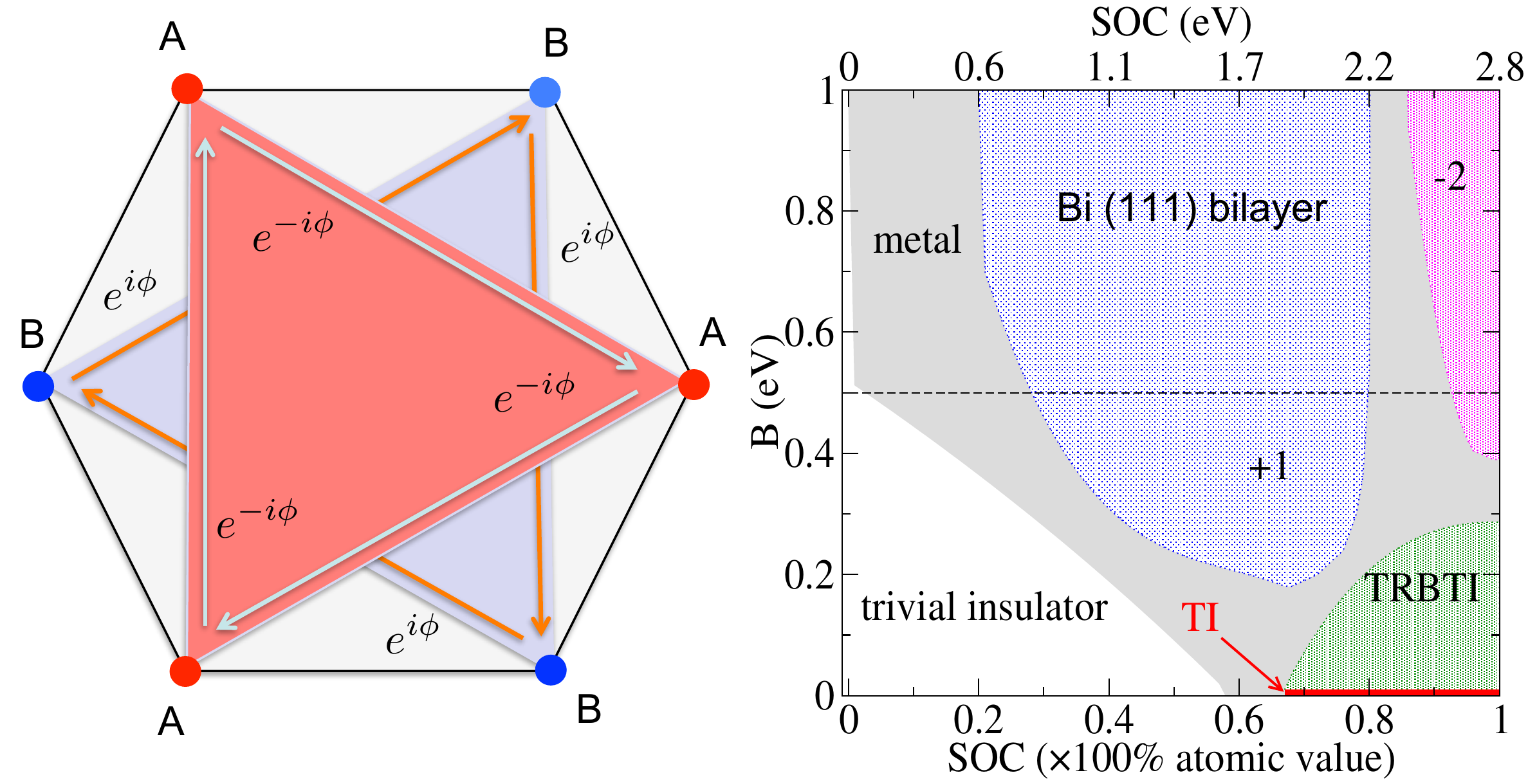}
\end{center}
\caption{\label{Fig-Haldane} Left: Real-space schematic representation of the Haldane model. 
Right: The phase diagram of the Bi(111) bilayer with respect to the strength of atomic SOC and 
magnitude of exchange field $B$. Numbers denote the Chern number in the quantum anomalous
Hall phase, ÒTIÓ stands for the topological insulator phase, while ÒTRBTIÓ stands for the 
time-reversal broken topological insulator phase. For more details see text and Ref.~\cite{Zhang}.}
\end{figure}

Historically, Thouless and co-workers~\cite{Thouless} were the first to demonstrate that Chern insulators can arise 
for periodic 2D solids exposed to an external magnetic field. Obtained in such a way Chern
insulator can be named the {\it quantum Hall insulator}, since the quantization of Hall conductance
in such Chern insulators was observed in measurements of the integer quantum Hall effect.
The quantum Hall insulators are to be distinguished from spontaneous Chern insulators, for which
the Chern insulator state is realized without external fields, and the breaking of time-reversal 
symmetry (necessary for non-zero Berry curvature) is the intrinsic property of the material due
to~e.g.~formation of local spin moments. We will refer to spontaneous Chern insulators as 
{\it quantum anomalous Hall insulators} (QAH insulators), since the quantization of conductance
in QAH insulators is observed by measuring the anomalous Hall effect (AHE) for which external
magnetic field plays a secondary role~\cite{Nagaosa}. The corresponding
effect is called the {\it QAH effect}.\iffindex{quantum anomalous Hall insulator}

Although the non-zero Chern number is due to non-trivial distribution of the Berry curvature in
$\bfk$-space, a fruitful analysis of QAH phases can be achieved in real space by considering 
various mechanisms of electron hopping and interactions on a lattice, and corresponding 
tight-binding Hamiltonians. The first lattice model for a QAH insulator was given by Haldane~\cite{Haldane}.
The Hamiltonian on the honeycomb lattice within the spinless {\it Haldane model} looks very simple:\iffindex{Haldane model}
\begin{equation}
H = t_1\sum_{\langle i,j\rangle }c_i^{\dagger}c_j + t_2\sum_{\langle\langle i,j\rangle \rangle}
e^{-i\sigma\phi}c_i^{\dagger}c_j,
\end{equation}
where the first term corresponds to the hopping between the nearest neighbors, while the second
term corresponds to the hopping between the next-nearest neighbors. The key feature  of the Haldane 
model is that the next-nearest neighbor hopping $t_2$ acquires a complex phase $e^{-i\sigma\phi}$,
where $\sigma=+1$ for the hopping on the A-sublattice, while $\sigma=-1$ for the hopping 
within the B-sublattice. The effect of aquiring a complex phase during electron hopping can be seen as
a result of a fictitious magnetic field with the vector potential $\bfA(\bfr)$: $e^{-i\phi}=
e^{-i\int d\bfr \bfA(\bfr)}$, where the integral in taken along the shortest path
which connect the next-nearest neighbor sites. As the electron completes a closed path
when hopping on the corresponding sublattice (see triangles in~Fig.~\ref{Fig-Haldane}), it accumulates
a phase which is proportional to the flux of the magnetic field through the corresponding triangle,
in analogy to the AB-effect we considered before.\footnote{The AB-effect with magnetic field opposite for electrons of
different spin, and not sublattice, will re-appear again in the context of the topological Hall effect.} 
Since this phase is opposite for electrons of 
two sublattices, the total field acting on electrons averages to zero within the unit cell.\footnote{That is
why the Haldane model is often called a model for a quantum Hall effect with zero magnetic
field.} Haldane showed by Fourier transforming the lattice Hamiltonian to the $\bfk$-space 
that the Chern number of this model equals $+1$ for $-\pi<\phi<0$,
and  $-1$ for $0<\phi<\pi$. The point $\phi=0$ thus gives rise to Weyl points in $(\bfk,\phi)$-space. Conceptually, the suggestion of his model by Haldane in 1988 
stands at the origin of the tremendous advances in topological condensed matter physics which followed.

One of the reasons for this is that the mechanism which gives non-zero Chern number within
the Haldane model can be realized in actual materials, with intrinsic spin-orbit interaction (SOI) (discussed in detail by Gustav Bihlmayer in manuscript A10 of this book) playing 
the role of the source of ``fictitious" magnetic field which provides non-trivial band
topology. To briefly demonstrate how this comes about we focus here on one of the many possible
examples considered by now in the literature~\cite{Bernevig}. Namely, we will consider the $p_z$ orbitals on a strongly buckled
honeycomb lattice of space-inversion symmetric (111) bilayer of Bismuth, 
see Fig.~\ref{Bi-bilayer}. The nearest-neighbor 
tight-binding multi-orbital lattice Hamiltonian of this system reads:
\begin{equation}\label{Bi}
H\ =\ \sum_{ij}t_{ij}c^\dagger_ic_j + \sum_ic_i^\dagger(\varepsilon_i \mathbb{I}+
B\sigma_z)c_i\ +H_{\text{SOC}},
\end{equation}
where the first term is the kinetic nearest-neighbor hopping between generally different multiple 
$s$ and $p$ (and $d$ or $f$ in transition and rare-earth metals) orbitals. The second term stands 
for an orbital on-site energy $\varepsilon_i$ and the interaction with the Zeeman
exchange field $B$ directed along the $z$-axis, with $\mathbb{I}$ ($\sigma_z$) as the identity (Pauli) matrix.
The third term in Hamiltonian~(\ref{Bi}) is the on-site SOI Hamiltonian.
Without the presence of $B$ the system has time-reversal symmetry and its bands are degenerate 
in spin throughout the whole BZ. We use the exchange field to break the time-reversal symmetry
and induce a non-zero QAH effect. To identify different origins of the Chern insulator phase, the 
spin-orbit interaction is further decomposed into spin-conserving and spin-flip parts:
\begin{equation}\label{SOI}
H_{\text{SOI}}\ =\ \xi\mathbf{l}\cdot\mathbf{s}=\xi l_zs_z+\xi(l^+s^-+l^-s^+)/2,
\end{equation}
where $\mathbf{l}$ ($\mathbf{s}$) is the orbital (spin) angular momentum operator, and $\xi$ is the
atomic SOI strength. Since in this work we choose the direction of the spin-polarization
to be aligned along $z$ direction, the spin conserving part of the SOI, $\xi l_zs_z$,
couples \{$p_x,p_y$\} orbitals, while the spin-flip part of Eq.~(\ref{Eq2}), $\xi(l^+s^-+l^-s^+)/2$, couples $p_z$ and
\{$p_x,p_y$\} orbitals via a flip of spin and a $\pm 1$ change in the orbital quantum number.

We concentrate here only on the case of the $p_z$ orbitals present around the Fermi energy, with other
states being pushed away much higher in energy. This case is particularly relevant for graphene
physics. For the values of the spin-orbit strength and hopping parameters we choose those of Bi from Ref.~\cite{Zhang}. First, consider the case when only spin-conserving SOI is present. On a buckled honeycomb lattice,
such as Bi bilayer or silicine, $p_z$ orbitals can hybridize directly with the \{$p_x,p_y$\} orbitals on 
the neighbring site, and complex hoppings within the sublattice can be induced via the spin-conserving
part of SOC which acts between $p_x$ and $p_y$ states. As illustrated with a sketch in 
Fig.~\ref{Bi-bilayer}, in this mechanism
the corresponding virtual transitions read:\,$
\ket{p_z^\text{A}\uparrow}\overset{t_{\text{NN}}}{\rightarrow}\ket{p_{x,y}^\text{B}\uparrow}
\overset{\xi l_zs_z}{\rightarrow}\ket{p_{x,y}^\text{B}\uparrow}
\overset{t_\text{NN}}{\rightarrow}\ket{p_z^\text{A}\uparrow} $,
where $t_{\text{NN}}$ indicates the direct hopping between $p_z$ and $p_{x,y}$ orbitals on the neighboring sites, while
superscripts A and B denote the nearest neighbor atomic sites in sublattice A  and B. The SOI here 
acts as a magnetic field which is responsible for the generation of phase $\phi$ within the Haldane 
model. If we consider only one spin, we can indeed show that effective SOI induced next-nearest-neighbor (NNN) hopping leads to the opening of the gap at the Fermi energy (of the size $\Delta_1$ in Fig.~\ref{Bi-bilayer}(a)) and the Chern number 
of the spin-up bands acquires a value of $+1$.  Since the SOI coefficients are complex conjugate for spin-up and spin-down electrons, the $\phi$ of the complex hoppings for spin-down electrons
is opposite in sign to spin-up electrons and same holds for the Chern numbers. Since the bands 
are spin-degenerate with $B=0$ 
(we apply as small exchange field in Fig.~\ref{Bi-bilayer} to artificially separate
bands of opposite spin for visibility), this results in a zero total Chern number, and 
the system resides in a {\it topological insulator phase} (see here also manuscript A10 by Gustav Bihlmayer).\iffindex{topological insulator} 

On-site spin-flip SOI can give rise to complex next nearest neighbor hopping too, even if there is no direct
hybridization between $p_z$ and \{$p_x,p_y$\} orbitals, Fig.~\ref{Bi-bilayer}(b).
In this case, the corresponding virtual transitions are: $
\ket{p_z^\text{A}\uparrow}\overset{\xi_\text{flip}}{\rightarrow}\ket{p_{x,y}^\text{A}\downarrow}\overset{t_{\text{NN}}}{\rightarrow}\ket{p_{x,y}^\text{B}\downarrow}\overset{t_\text{NN}}{\rightarrow}\ket{p_{x,y}^\text{A}\downarrow}\overset{\xi_\text{flip}}{\rightarrow}\ket{p_z^\text{A}\uparrow} $ 
where $\xi_\text{flip}=\xi(l^+s^-+l^-s^+)/2$, and $t_{\text{NN}}$ stands for the direct hybridization between $p_{x,y}$ orbitals on neighboring A and B sites. The corresponding NNN hopping is again
between electrons of the same spin, owing to two spin-flip processes which take place in between.
In analogy to the case with spin-conserving SOI considered previously, the effective hoppings within A and B sublattices for fixed spin are of opposite sign and the resulting gap $\Delta_2$ which opens due to latter
virtual transitions is again topologically nontrivial. The resulting non-zero Chern numbers of the
degenerate without $B$ spin-up and spin-down bands are exactly the same as previously.
As in the previous case, the coupling between the spin-up and spin-down $p_z$ bands does
not occur, and without an exchange field, Fig.~\ref{Bi-bilayer}(b) corresponds exactly to the 
topological insulator phase in graphene~\cite{Kane}.

\begin{figure}[t!]
\begin{center}
\includegraphics[width=0.85\textwidth]{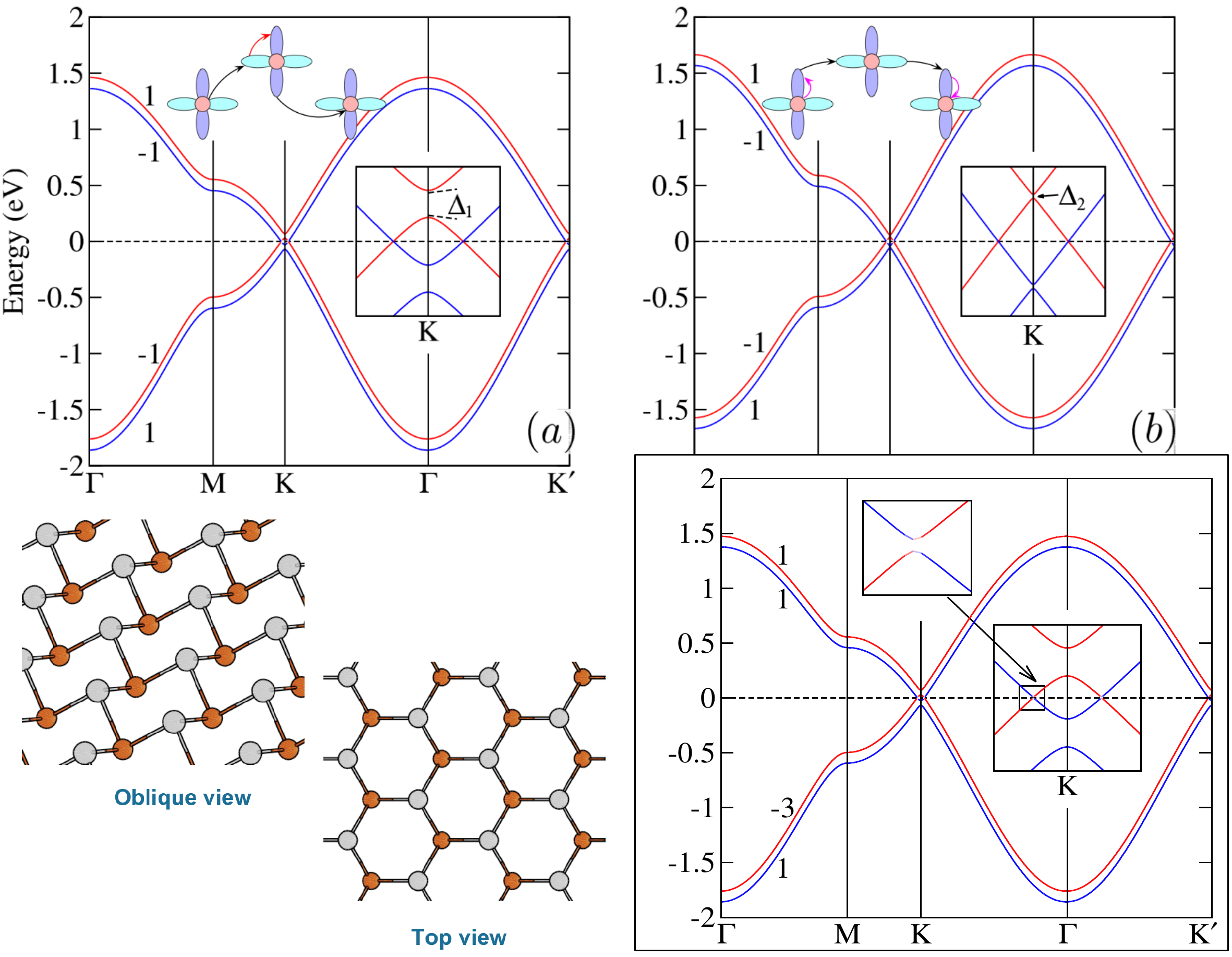}
\end{center}
\caption{\label{Bi-bilayer} Topological analysis of $p_z$ bands in a buckled honeycomb bilayer. A small exchange field has been applied in all cases, yielding a QAH state in lower right figure. Without an exchange field, systems in (a) and (b) would be in a topological insulator state. Left up and right up figures correspond to the case with only spin-conserving SOC, and only spin-flip SOC included, while the full SOC is considered in the lower right figure.  Red (blue) stands for the spin-down (spin-up) states. Dashed horizontal lines indicate the Fermi energy. In the case of isolated bands, numbers denote the Chern number for each individual band, while for overlapping bands, numbers stand for the Chern numbers of nonhybridizing spin-up (red) and spin-down (blue) bands. Insets display the electronic structure of the Dirac point at the Fermi energy, and sketches illustrate different channels for complex nearest-neighbor hopping [red circles denote $p_z$ orbitals, while $p_{x,y}$ orbitals are indicated by blue ellipsoids; black (red) arrows depicts the nearest-neighbor hoppings (SOC hybridization), respectively]. Lower left: top and side view of the Bi(111) bilayer. For more details see text and Ref.~\cite{Zhang}.}
\end{figure}

How do we make Chern insulators out of systems above? In principle, if we could apply a very
strong exchange field which would shift the bands of a certain spin very high up in energy,
we would readily obtain a QAH insulator. In real materials this is however seldomly achievable,
since the magnitudes of typical exchange fields are normally smaller than the typical band width.
Thus, the only way would be in breaking the time-reversal symmetry with a finite $B$, and ensuring
that a topologically non-trivial band gap opens at the Fermi energy where bands of opposite spin
meet. For example, we could start with a situation depicted in Fig.~\ref{Bi-bilayer}(a) with a small
$B$, and add an admixture of the spin-flip SOI to the Hamiltonian. This will open a gap at the 
points where the spin-up and spin-down bands were degenerate, see Fig.~\ref{Bi-bilayer}. In
the vicinity of such a point the distribution of spin becomes non-trivial, as we can see in Fig.~\ref{Bi-bilayer}, namely, e.g. at the K-point the spin-distribution of the occupied band has a skyrmion
structure. At this point, we can relate the distribution of spin to the distribution of vector $\bfd$ and, according to the two-band analysis presented above, we can explain the fact that the Chern
number of the occupied band changes. The spin distribution of the corresponding conduction band
is also a skyrmion, but with an opposite winding number, which results in the opposite change of 
the Chern number of this band. What we have just achieved is the exchange of the Chern number
between the bands of opposite spin at the points at the Fermi energy where bands of opposite
spin hybridize. We call such points {\it spin-mixing points}. In this particular example the spin-mixing
point is the Weyl point in the space of $(k_x,k_y,\xi_\text{flip})$, where $\xi_\text{flip}$ is the strength
of the spin-flip SOI. We have two spin-mixing Weyl points in our 3D space: (K, $\xi_\text{flip}=0$)
and (K$'$, $\xi_\text{flip}=0$). The change in the Chern number upon going through these
Weyl points can be computed based on the dispersion of the bands, and the total Chern number of all occupied states can be calculated to be $-2$ for a realistic situation of
$\xi_\text{flip}>0$. We have thus achieved a QAH state. 

We remark here, that Hamiltonians of
real materials can be very complicated with many states present at the Fermi energy and various 
structural, spin-orbit and magnetic effects taking place. The phase diagrams of such materials as
a function of parameters in the Hamiltonian can be studied from first principles methods. See for
example the phase diagram of Bi(111) bilayer as a function of an exchange field and SOI strength
calculated from {\it ab initio} in Fig.~\ref{Fig-Haldane}.  

We would like to comment on the relation between Chern numbers and transport properties
of 2D insulators. We have learned by now that a Chern insulator has a quantized transverse charge
conductance, proportional to the value of the Chern number. This means that e.g. for the case
of Fig.~\ref{Bi-bilayer}(a) the transverse charge conductance vanishes. The carriers of each spin
separately, however, possess a quantized charge conductance. In an applied electric field the
carriers of opposite spin will move in opposite directions, which will lead to the generation
of the transverse {\it spin current}. Thus obtained spin conductance is quantized since the conductance
for each spin is quantized, and it is proportional to the difference between the Chern numbers 
for each spin. This number is called the {\it spin Chern number}. The system of Fig.~\ref{Bi-bilayer}(a)
is a simple example of a {\it quantum spin Hall} (QSH) insulator, or, equivalently, a 2D topological 
insulator. The concept of the spin Chern number can be generalized to the cases where spin
is not conserved, and even to the case of broken time-reversal symmetry. For
2D insulators with time-inversion the spin Chern number can be used alternatively to the $\mathbb{Z}_2$ index in order to classify topological phases~\cite{Bernevig}. 

In metals, the Chern number can be formally calculated but it is not quantized, since for bands which
cross the Fermi energy the integral of the Berry curvature goes only along the patches of the BZ where
the band is occupied. Nevertheless, the so-calculated integral of the Berry curvature over all occupied
states gives the value of the {\it intrinsic} anomalous Hall effect (AHE).\iffindex{anomalous Hall effect} In contrast to insulators, the presence of Fermi surface in metals also leads to promotion of the  Hall current which comes from impurity scattering $-$ this is the so-called {\it extrinsic} AHE. The relation between the magnitudes of the intrinsic and extrinsic currents strongly depends on the details of the electronic
structure and disorder~\cite{Nagaosa}. The theory of impurity scattering will be considered in detail by  Phivos Mavropoulos in manuscript A5
of this book. In analogy to the relation between the Chern and spin Chern numbers, if the spin
is conserved, the anomalous Hall conductivity can be decomposed into a sum of contributions 
coming from spin-up and spin-down bands. The difference between the two is proportional to the
value of the spin Hall conductivity, that is, it is related to the magnitude of the transverse spin 
current caused by an electric field. This effect is called the {\it spin Hall effect} (SHE).\iffindex{spin Hall effect}

\subsection{Dynamics of wavepackets in solids}
A very powerful approach to study the properties of solids lies in rewriting the problem in terms of so-called wavepackets. The general description of the electron dynamics in external fields and
perturbations can be rigorously provided referring to semiclassical dynamics of such wavepackets. The wavepackets are obtained by a convolution of Bloch states with the envelope
function which is centered around a certain $\bfk$-vector. The wavepackets are made such that they are localized both
in reciprocal and real space, and the evolution of their center of mass in both spaces can be
directly related to the transport characteristics of a solid. The quantum mechanical description
of wavepackets is an intricate science on its own, and here, we will only provide intuitive arguments
which can be used to understand how the exact equations of wavepacket motion are obtained~\cite{Xiao}.\\

{\it \textbf{Uniform electric field}.} \\
For simplicity, let us first consider the effect of an electric field $\bfE$ present in a solid. Let us say
that without the field the unperturbed Hamiltonian of the system looks like
\begin{equation}
H(t=0) = \frac{\bfp^2}{2} + V(\bfr),
\end{equation}
with corresponding Bloch vectors $\bfq$ and the crystal momentum representation $H_{\bfq}$ of the Hamiltonian. 
To model the effect of the uniform electric field, we apply 
a constant in space, but varying in time vector potential $\bfA(t)$ such that 
$-\frac{\partial \bfA(t)}{\partial t}=\bfE$. This modifies the corresponding
lattice Hamiltonian as follows:
\begin{equation}
H(t>0) = \frac{1}{2}\left(\bfp +\bfA(t)\right)^2 + V(\bfr).
\end{equation}
Since the constant in space vector potential does not break the periodicity of the crystal, it
cannot couple the unperturbed wavefunctions with different values of $\bfq$ and it changes the
energy of the states with an overall constant, which can be ignored as we shall see later. Therefore,
once we are looking at a state labeled with a certain value of $\bfq$, during the evolution
of this state $\dot{\bfq}=0$. Nevertheless, we can also number our state in terms of the 
$\bfk$ vector, which is the ``proper" Bloch vector of the Hamiltonian $H_{\bfk}(t>0)$, and which
is called {\it gauge-invariant momentum}. 
The relation between the $\bfk$ and $\bfq$ reads as follows:
\begin{equation}
\bfk = \bfq + \bfA(t) = \bfk(t),
\end{equation}
that is, the wavefunction at $\bfk(t)$ which solves the Schr\"odinger equation for $H(t)$, is
identical to the wavefunction which solves the Schr\"odinger equation for $H(t=0)$ but at
a wavevector 
$\bfq = \bfk - \bfA(t)$. 
From the latter equation, it follows that
\begin{equation}
\dot{\bfk} = -\bfE.
\end{equation}
In order to employ the expression for the velocity of a certain state, we write the time-dependence
of the Hamiltonian as follows: $H_{\bfk}^t=H_{\bfk(t)}$, which leads to $\partial_t H_{\bfk}^t=
\partial_{\bfk}H_{\bfk}\cdot\partial_t\bfk(t) = -\bfE\cdot\partial_{\bfk}H_{\bfk}$. Substituting 
the latter into the expression for the velocity (\ref{106}) with $t$ playing the role of $\lambda$, we get:
\begin{equation}
\dot{\bfr}:=v_{n\bfk}=\partial_{\bfk}\varepsilon_{n\bfk}-\bfE\times\boldsymbol\Omega_n(\bfk).
\end{equation} 
From these equations we can easily understand the Berry phase origin of the intrinsic anomalous
Hall effect in ferromagnets, as well as the expression for the Hall conductance in terms of the Berry
curvature given by Eq.~(\ref{2D-Hall}), taking into account that the contribution from the group
velocity when integrated over the BZ, vanishes. \\

{\it \textbf{Uniform magnetic field}.}\\
 Let us assume now a situation in which we have applied an external
uniform magnetic field $\bfB={\rm curl}\bfA(\bfr)$.  This situation is analogous to the previously
considered case only with $\bfA(\bfr)$ replacing the $\bfA(t)$, but the way of treating the two
situations is quite different. Namely, generally speaking, the vector potential $\bfA$ breaks the 
periodicity of the lattice. This obstacle can be overcome by assuming that the vector potential varies
very slowly in space, so that locally at a given point $\bfR$ in space the periodicity of the lattice
is preserved and the ``local" Bloch momentum $\bfk$ is well-defined. Seen from point $\bfR$, the
presence of the vector potential is then just a constant shift of the momentum operator, analogously to
the case of the electric field. The time-dependent process associated with this assumption is the
propagation of an electron wavepacket, with the center in real space at $\bfR(t)$, through a slowly 
varying medium with the 
Hamiltonian $H=H(\bfk(\bfR(t))$, where the gauge-invariant momentum $\bfk$ is given by:
\begin{equation}  
\bfk = \bfq + \bfA(\bfR) = \bfk(t)=\bfk(\bfR(t)).
\end{equation}
Following the same logic as previously, that is that $\dot{\bfq}=0$, we derive the equation of
propagation in the reciprocal space:
\begin{equation} 
\dot{\bfk} = \bfB\times\dot{\bfR}=\bfOmega_{\bfR}\times\dot{\bfR},
\end{equation}
where $\bfOmega_{\bfR}$ is real space Berry curvature (\ref{232}), which corresponds to the magnetic
field, as we saw from the consideration of the AB-effect.
On the other hand, since $\partial_t H_{\bfk}^t=
\partial_{\bfk}H_{\bfk}\cdot\dot{\bfk}$, the equation of motion of the wavepacket generalizes 
to
\begin{equation}
\dot{\bfR}=\partial_{\bfk}\varepsilon_{n\bfk}+\boldsymbol\Omega_n(\bfk)\times\dot{\bfk}.
\end{equation} 

{\it \textbf{Explicit dependence on $\bfR$}.} \\
Imagine now that the dependence of the local Hamiltonian 
on  the slowly varying
spatial coordinate $\bfR$ is not only via the vector potential, but also via the crystal potential
$V(\bfR)$,
or ($\bfR$-dependent) exchange coupling of the spin of a propagating electron to the 
($\bfR$-dependent) magnetic texture (the case we will
consider in detail in the next section). Namely, let us suppose that the Hamiltonian assumes a 
more general dependence $H=H(\bfk(\bfR(t)),\bfR(t))$. In this case the time-derivative of the
Hamiltonian
\begin{equation}
\partial_t H(\bfk(\bfR(t)),\bfR(t)) = \partial_{\bfk}H_{\bfk\bfR}\cdot\dot{\bfk} +
 \partial_{\bfR}H_{\bfk\bfR}\cdot\dot{\bfR}.
\end{equation}
This leads to the fact that the velocity of a wavepacket through the $\bfR$-texture acquires an
additional contribution due to the  {\it mixed Berry curvature} $\Omega_{\bfk\bfR}$,
which can be expressed in terms of the derivatives of $u_{\bfk\bfR}$ with respect to $\bfk$ and
with respect to $\bfR$. This can be intuitively understood by looking at Eq.~(\ref{106}), into which both of these derivatives will enter upon the time evolution of the states. We previously defined the mixed Berry curvature in the context of electric
polarization, for which $\bfR$ was replaced with $\lambda$.\\

{\it \textbf{General case.}}\\
 While the hand-waving arguments we provided to derive the expression for the
velocity of a wavepacket propagating through an $\bfR$-texture are qualitatively correct, the rigorous
quantum mechanical derivation of the equations of motion of the center of a wavepacket in the $(\bfR,\bfk,t)$ phase space has been given by Sundaram and Niu in Ref.~\cite{Sundaram}. If we have an explicit 
time-dependence of the Hamiltonian (e.g.~via a moving magnetic texture), we should assume the 
general dependence of the Hamiltonian $H = H(\bfk,\bfR,t)$, given that the texture varies 
very slowly in space and in time. In the latter expression for the Hamiltonian, $\bfk$ stands for
the ``local" Bloch vector, which is a good quantum number under the assumption that locally
around the point $\bfR$ in space the texture can be approximated as a constant and the lattice
periodicity is preserved. In terms of preceeding subsections $\bfk$ plays the role of $\bfq$.
Correspondingly, all wavefunctions acquire additional dependence on
$\bfR$ and $t$: $u_{n\bfk}\longrightarrow u^t_{n\bfk\bfR}$. In accordance to this additional 
dependence, we can introduce the following gauge potentials:
\begin{equation}\label{potentials}
\calA_t^n = i\Braket{u^t_{n\bfk\bfR}|\partial_t u^t_{n\bfk\bfR}},\quad
\calA^n_{\bfk} = i\Braket{u^t_{n\bfk\bfR}|\partial_{\bfk} u^t_{n\bfk\bfR}},\quad
\calA^n_{\bfR} = i\Braket{u^t_{n\bfk\bfR}|\partial_{\bfR} u^t_{n\bfk\bfR}},
\end{equation}
which can be all seen as one gauge potential $\calA^n$ on the $(\bfR,\bfk,t)$-manifold with components 
given above. The corresponding curvature of our phase space $\Omega_{ij}^n=\partial_i\calA^n_j - \partial_j\calA^n_i$ has then corresponding components:
\begin{equation}\label{curvatures}
\Omega^n_{\bfk_i\bfk_j}=-2{\rm Im}\Braket{\partial_{\bfk_i} u^t_{n\bfk\bfR}|
     \partial_{\bfk_j} u^t_{n\mathbf{k}\bfR}},\quad
\Omega^n_{\bfR_i\bfR_j}=-2{\rm Im}\Braket{\partial_{\bfR_i} u^t_{n\bfk\bfR}|
     \partial_{\bfR_j} u^t_{n\mathbf{k}\bfR}},
\end{equation}
and the mixed curvature:
\begin{equation}\label{curvature}
\Omega^n_{\bfk_i\bfR_j}=-2{\rm Im}\Braket{\partial_{\bfk_i} u^t_{n\bfk\bfR}|
     \partial_{\bfR_j} u^t_{n\mathbf{k}\bfR}}.
\end{equation}
The expressions for the time-involving components of the curvature can be written 
analogously. The equations of motion of the center of the wavepacket which is centered at 
points $\bfR$ and $\bfk$ in real and reciprocal space, respectively, then read:
\begin{eqnarray}\label{132}
\dot{\bfR} = \partial_{\bfk}\mathcal{E}^t_{n\bfR\bfk} - \left(  \Omega^n_{\bfk\bfR}\cdot\dot{\bfR}  
+ \Omega^n_{\bfk\bfk}\cdot\dot{\bfk} \right) - \Omega^n_{\bfk t}, \\ \label{133}
\dot{\bfk} = -\partial_{\bfR}\mathcal{E}^t_{n\bfR\bfk} + \left(  \Omega^n_{\bfR\bfR}\cdot\dot{\bfR}  
+ \Omega^n_{\bfR\bfk}\cdot\dot{\bfk} \right) - \Omega^n_{\bfR t}. 
\end{eqnarray}
Note that the band energy $\varepsilon^t_{n\bfk\bfR}$ acquires an additional contribution:
\begin{equation}\label{ENERGY}
\mathcal{E}^t_{n\bfR\bfk} = \varepsilon^t_{n\bfk\bfR}+ \delta\varepsilon^t_{n\bfk\bfR} = \varepsilon^t_{n\bfk\bfR} - {\rm Im}
\sum_i\Braket{\partial_{\bfR_i} u^t_{n\bfk\bfR}|\varepsilon^t_{n\bfk\bfR} - H(\bfk,\bfR,t)|\partial_{\bfk_i} u^t_{n\bfk\bfR}}.
\end{equation}
In the context of situations considered previously, it is clear that the $\Omega^n_{\bfR\bfR}$ curvature
plays the role of the real-space magnetic field, while the $\Omega^n_{\bfk\bfk}$ is the curvature in
reciprocal space which gives rise to the anomalous Hall effect. When the system is subject to
constant electric and magnetic fields, the mixed Berry curvature is zero, and equations of motion
for the wavepackets are reduced to:
\begin{eqnarray}
\dot{\bfR} &=& \partial_{\bfk}\mathcal{E}_{n\bfk} - \dot{\bfk}\times\bfOmega^n_{\bfk\bfk}\\
\dot{\bfk} &=& -\bfE - \dot{\bfR}\times\bfB,
\end{eqnarray}
where $\bfk$ stands now for gauge-invariant momentum from before: $\bfk=\bfq+\bfA(\bfR)$. If we use now that $\partial_{\bfR}=(\partial_{\bfR}\mathcal{A})\partial_{\bfk}$, we can 
write down the contribution to the energy in terms of $\bfk$-derivatives only:
\begin{equation}\label{Eorb}
\mathcal{E}_{n\bfk} = \varepsilon_{n\bfk} - \bfB\cdot{\rm Im}
\Braket{\partial_{\bfk} u_{n\bfk}|\times[\varepsilon_{n\bfk} - H(\bfk,t)]|\partial_{\bfk} u_{n\bfk}}=
\varepsilon_{n\bfk} -\bfB\cdot\bfm(\bfk).
\end{equation}
In the last equation, $\bfm(\bfk)$ is given by:
\begin{equation}
\bfm(\bfk) = {\rm Im}\Braket{\partial_{\bfk} u_{n\bfk}|\times[\varepsilon_{n\bfk} - H(\bfk,t)]|\partial_{\bfk} u_{n\bfk}},
\end{equation}
and it stands for the orbital moment of the wavepacket corresponding to the ``internal" degree
of freedom with respect to the rotation around its own axis. The last term in Eq.~(\ref{Eorb}) thus
corresponds to the interaction of the magnetic field with the orbital moment of the wavepacket
due to rotation around its own axis.
\\

{\it \textbf{Geometrical meaning.}}\\
Let us for simplicity drop the explicit time dependence in Eqs.~(\ref{132}-\ref{133}). Then the equations which govern the
dynamics in the phase space $\bfx=(\bfk,\bfR)$ can be written in the following form:
\begin{equation}\label{HAM}
(\Omega^n - J)\dot{\bfx} = \partial_{\bfx}\mathcal{E}_n \quad \Longleftrightarrow \quad
\omega^n_{\alpha\beta}\dot{\bfx}_{\beta} = \partial_{\bfx_{\alpha}}\mathcal{E}_n
\end{equation}
where for simplicity we dropped $\bfR$ and $\bfk$ indices, $\omega^n =(\Omega^n-J)$ and $J= \begin{pmatrix}
		0 & I \\
		-I & 0
	\end{pmatrix}$.
These equations are nothing else but the equations of the classical Hamiltonian dynamics written 
in terms of {\it non-canonical} variables $\bfx$ for the Hamiltonian of the form $h(\bfx)=\mathcal{E}_n(\bfx)$.
If we introduce the Poisson bracket between two functions $f(\bfx)$ and $g(\bfx)$ as follows:
\begin{equation}
\{f,g\} := \partial_{\bfx}f^T \cdot \left(\omega^{n}\right)^{-1} \cdot \partial_{\bfx}g, 
\end{equation}
the Hamilton equations acquire the standard form:
\begin{equation}
\dot{\bfx} = \{  \bfx, h \}.
\end{equation}
The fact that the position and momentum of the wavepacket are non-canonical variables comes
with a price. Namely, the canonical Hamilton equations, obtained with $\omega=-J$, satisfy the property called the {\it Liouville's theorem}, which states that during the Hamiltonian dynamics the 
volume of a certain region in phase space, say, $d\bfk d\bfR$, is preserved. It is very easy to check,
that, taking initially an infinitesimal volume $d\bfk d\bfR$ and evoluting it in accordance to (\ref{HAM})
will violate the Liouville's theorem. The reason for this lies in the fact that $\bfk$ and $\bfR$ are 
non-canonical variables, when the curvature form is non-zero. Mathematically speaking, we endow 
our phase space with a non-canonical simplectic two-form $\omega^n=(1/2)\omega^n_{\alpha\beta}d\bfk_{\alpha}\wedge d\bfR_{\beta}$,  which influences the measure of our $\bfx$-space. It is straightforward to see that the infinitesimal volume $d\bfk d\bfR$ multiplied with the function,
called {\it modified density of states},  
\begin{equation}\label{D-den}
D^n(\bfx)=\frac{1}{(2\pi)^d}\sqrt{\det \omega^n}
\end{equation}
satisfies the Liouville's theorem. 
In case of canonical variables the 
modified density of states (DOS),
$D (\bfx) = 1/(2\pi)^d$ (multiplied with the Fermi distribution function). 
Let us take a detailed look at the explicit expression for the case of the (generally 
$\bfR$-dependent) magnetic field in the system. The Poisson brackets between the $\bfR$ and $\bfk$ coordinates in this case read:
\begin{equation} 
\{ \bfR_i,\bfR_j\} = \frac{\varepsilon_{ijl}\bfOmega^n_{\bfk\bfk,l}}{1+\bfB\cdot\bfOmega^n_{\bfk\bfk}},\quad
\{ \bfk_i,\bfk_j\} = -\frac{\varepsilon_{ijl}\bfB_l}{1+\bfB\cdot\bfOmega^n_{\bfk\bfk}},\quad
\{ \bfR_i,\bfk_j\} = \frac{\delta_{ij} +\bfB_i\bfOmega^n_{\bfk\bfk,j}}{1+\bfB\cdot\bfOmega^n_{\bfk\bfk}},
\end{equation}
while the determinant of $\omega$ gives:
\begin{equation}
\sqrt{\det \omega^n} = 1 + \bfB\cdot\bfOmega_{\bfk\bfk} \quad \Longrightarrow \quad D^n(\bfx)=\frac{1}{(2\pi)^d}(1 + \bfB\cdot\bfOmega_{\bfk\bfk})
\end{equation}
The volume element $D^n(\bfx)d\bfk d\bfR$ has constant density of quantum states and it has to be used for 
computing the expectation values of the observables obtained by an integration over the phase space. We have to
understand that, of course, coordinates $\bfk$ and $\bfR$ can be made canonical, although possibly less appealing
physically. For such canonical variables, say, $\tilde{\bfk}$ and $\tilde{\bfR}$, the modified DOS reads 
$D^n(\tilde{\bfx})=1/(2\pi)^d$ and the very same expectation values can be obtained by a direct evaluation of the integral with the
volume element measure $d\tilde{\bfk} d\tilde{\bfR}$.  
The situation here is analogous to switching between integrals
of a function in $\mathbb{R}^3$ performed for example in cartesian and spherical coordinates. 
The modified DOS for the non-canonical variables is
thus analogous to the Jacobian of the transformation between cartesian and spherical coordinates.
Below we briefly outline the consequences of the non-canonicity of the $\bfR$ and $\bfk$
coordinates for a selected number of physical properties of the system.\\

{\it \textbf{Fermi volume.}}\\
The change in the density of electronic states in the $(\bfR,\bfk)$ space inevitably changes the
expectation values of quantum operators which are obtained as integrals over $(\bfR,\bfk)$ space. Consider 
the Fermi volume which a given number of electrons occupies. The number of occupied electrons, $N_e$, is given by:
\begin{equation}
N_e = \int_{-\infty}^{E_F} \frac{d\bfk}{(2\pi)^d} (1 + \bfB\cdot\bfOmega_{\bfk\bfk}).
\end{equation}
In case of a small magnetic field, 
keeping the number of electrons in an insulator constant, we come to the change in the Fermi volume by:
\begin{equation}
\Delta V_F = -\int_{\rm BZ} \frac{d\bfk}{(2\pi)^d} \,\bfB\cdot\bfOmega_{\bfk\bfk}.
\end{equation}
In case of two dimensions, we therefore get that $\Delta V_F=\sigma_{xy}\cdot B$.\\

{\it \textbf{Quantum Hall conductance.}}\\
Thermodynamically, the change in the free energy of the system can be written as:
\begin{equation}\label{F-energy}
dF = -\bfM_L\,d\bfB - N_e\,d\mu -S\,dT,
\end{equation} 
where $\bfM_L$ is the magnetization induced by the magnetic field $\bfB$, $\mu$ is the chemical
potential, $S$ is entropy and $T$ is temperature.
Streda formula states that the Hall conductivity of a two-dimensional finite sample is the derivative 
of the electron number at a certain chemical potential and temperature with respect to an applied field:
\begin{equation}
\sigma_{xy} = -\left(  \partial N_e / \partial \bfB  \right)_{\mu,T}.
\end{equation}
Using the expression derived previously for $N_e$, we obtain the known expresion for the $\sigma_{xy}$
in therms of the Berry curvature in $\bfk$-space. The Streda formula can be understood intuitively
by thinking in terms of a time dependent rise of the magnetic field in some region, which will 
generate an electric field along the boundary of this region. This in turn will cause the ``leakage" of the
charge from the region due to the anomalous Hall effect.
\\

{\it \textbf{Orbital magnetization.}} \\
The correction to the density of states allows us also to derive an explicit expression for the orbital magnetization from (\ref{F-energy}), since it is defined as $\bfM_L = -(\partial F/\partial \bfB)_{\mu,T}$. The expression for $\bfM_L$ can be derived referring to the explicit expression for $F$ ($\beta=1/k_BT$, and we drop the summation over bands for simplicity):
\begin{equation}\label{FREE}
F = -\frac{1}{\beta(2\pi)^d}\int d\bfk\left(1 +\bfB\cdot\bfOmega_{\bfk\bfk}\right)
\ln \left(  1 + e^{-\beta(\mathcal{E}_{n\bfk}-\mu)} \right).
\end{equation}
By differentiating this expression with respect to the magnetic field, we obtain:
\begin{equation}
\bfM_L = \int_{\rm BZ}\frac{d\bfk}{(2\pi)^d}\,f(\bfk)\bfm(\bfk) + \int_{\rm BZ}
\frac{d\bfk}{(2\pi)^d}\,\bfOmega_{\bfk\bfk}\ln \left(  1 + e^{-\beta(\varepsilon_{n\bfk}-\mu)} \right),
\end{equation}
where $f$ is the Fermi occupation function. From this expression it is clear that two effects contribute to the total orbital magnetization in a 
solid. The first one comes from the $\bfm(\bfk)$ correction to the energy, which can be identified
with the orbital moment of a wavepacket as it rotates around its axis. The second contribution comes
from the center-of-mass motion of the wavepackets, and can be expressed in terms of the anomalous
Hall conductivity.  
This latter contribution is thus a direct consequence of the modification in the density
of states in the phase space due to its non-trivial structure as expressed in terms of the curvature. At zero temperature, the expression for the orbital magnetization thus reads:
\begin{equation}\label{M-orb-0T}
\bfM_L = \sum_n\int_{\rm BZ}\frac{d\bfk}{(2\pi)^d}\,f_{n\bfk}\left[\bfm_n(\bfk) + 
(\varepsilon_{n\bfk} - \mu)\bfOmega^n_{\bfk\bfk}\right].
\end{equation}
The expression for the  quantized Hall conductivity in an insulator can be rederived
using the Maxwell relation $ (\partial \bfM_L/\partial \mu)_{\bfB,T} = (\partial N_e/\partial \bfB)_{\mu,T}$. The latter expression suggests that in a Chern insulator the orbital magnetization
varies linearly with the chemical potential. 
The mechanism for this effect lies in the presence of metallic states at the
boundary of a Chern insulator.\iffindex{orbital magnetization}

\subsection{Topological Hall effect}

{\it \textbf{Emergent field and topological Hall effect}.}\\The term topological Hall effect (THE) is normally referred to in the context of a magnetic 
material which exhibits a spatial variation of the magnetization $\bfM(\bfr)$. The simplest
possible Hamiltonian in this case, which gives rise to non-trivial effects, reads:
\begin{equation}\label{3.4-1}
H = \frac{\bfp^2}{2} - J\bfsigma\cdot\bfM(\bfr),
\end{equation}
where $J$ is the coupling constant, $\bfsigma$ is the vector of Pauli matrices, and we shall 
assume that the magnitude of the magnetization is constant,  that is, 
$\bfM(\bfr)=M{\bfn}(\bfr), |\bfn(\bfr)|=1$. How do we attack the problem of finding
the transport properties of such a system? The brute force method suggests that we 
take a finite sample, diagonalize the Hamiltonian, find
the spectrum and wavefunctions, and calculate the required properties. If the
magnetization texture exhibits a periodic lattice structure, we can even apply the Bloch
apparatus to arrive at the spectrum $E_{n\bfkappa}$ and wavefunctions $\Psi_{n\bfkappa}$,
where $\bfkappa$ is the Bloch vector from the reciprocal space.   
How does the Berry phase enter into this picture?

Practically, the Berry phase viewpoint at the problem is motivated by the consideration that
very often the typical scale of the variation of the texture in real space is very large, which
makes the brute force approach cumbersome. In this way we assume that locally around
a point $\bfR$ in real space the magnetization direction is constant, resulting in a set of
eigenvalues and wavefunctions of the ``local" Hamiltonian, $\varepsilon_{n\bfR}$ and 
$\psi_{n\bfR}$, respectively. The properties of the system are then related to the dynamics
of electrons as they move through the texture, or, in other terms, as parameter $\bfR$
is varied $-$ the Hamiltonian is then seen as a parametrized Hamiltonian $H(\bfR)$. This 
is the typical setup in which Berry phase physics arises. 

Let us analyze the situation of Eq.~(\ref{3.4-1}) from the standpoint of sections 2.4 and 3.3. We introduce
a mapping (\ref{chi}), $\chi: \mathbb{R}^3 \rightarrow S^2$, which maps the direction of $\bfn$ at
point $\bfR\in\mathbb{R}^3$ to a point on $S^2$,
which can be ascribed angles $\theta$ and $\varphi$. This mapping is analogous to that given by (\ref{chi}),
where the $\bfk$-space is replaced with $\bfR$-space. We are now to study the {\it adiabatic} dynamics of a wavefunction which solves Eq.~(\ref{3.4-1})
as it follows a certain trajectory $\mathcal{C}$ in real space, or, equivalently, a corresponding
trajectory $\chi(\mathcal{C})$ on $S^2$, see Fig.~\ref{Sphere}. In spin space, at each point on $S^2$, as
in the case of sections 2.4 and 3.3, we have two eigenenergies 
$\varepsilon_{\uparrow(\downarrow)}=\mp JM$ (assuming for simplicity that kinetic 
energy is zero), and wavefunctions $\psi_{\uparrow(\downarrow)}(\bfR)$ which solve (\ref{3.4-1}) for  corresponding $\bfR$. The
condition of the adiabaticity, as we defined it previously, requires that the time scale of the
evolution of the wavefunction along $\mathcal{C}$ is much smaller than the typical time-scale of the transitions which cause spin-flip events. This corresponds to the situation of remaining within
the same spin-up or spin-down subband during the dynamics. In this case the situation considered 
here can be made conceptually and technically analogous to that we considered for Chern insulators in section 3.3 
by replacing $\bfR$ with $\bfk$. 

As we recall from section 2.4, the problem posed on $S^2$ accounts to two independent copies of the 
Dirac monopole with the charge $+1$ and $-1$ for spin-up and spin-down electrons, respectively. 
Each spin $\sigma$ endows $S^2$ with the Berry connection $\mathcal{A}^{\sigma}_{\bfn}$ defined on northern 
and southern hemispheres. The corresponding curvature $\Omega^{\sigma}_{\bfn}$ has components 
which correspond to the field of a magnetic monopole at the origin, according to (\ref{Sp-A}), and has
opposite sign for up and down electrons, so that summed up over both states it gives zero. 
Then the Berry phase $\gamma(\mathcal{C})$ an electron of certain $\sigma$
picks up when it travels along $\mathcal{C}$ can be computed from the corresponding Berry phase of
the path $\chi(\mathcal{C})$. The connection between the two Berry phases can be computed by
writing down the relation between the Berry curvatures in both spaces. Namely, $\Omega^{\sigma}_{\bfn}$
and 
\begin{equation}
\Omega^{\sigma}_{\bfR}\equiv\Omega^{ij,\sigma}_{\bfR\bfR}=-2\rm{Im}\braket{\partial_{\bfR_i}\psi_\sigma(\bfR)|\partial_{\bfR_j}\psi_\sigma(\bfR)}
\end{equation} 
are connected in a way similar to that we considered for Chern insulators, Eq.~(\ref{113}):
\begin{equation}\label{159}
\Omega^{ij,\sigma}_{\bfR}=\sigma\,\bfn\cdot\left(   \partial_{\bfR_i}\bfn \times \partial_{\bfR_j}\bfn \right)/2 =
\bfOmega^{\sigma}_{\bfn}\cdot\left( \partial_{\bfR_i}\bfM \times \partial_{\bfR_j}\bfM   \right).
\end{equation}
According to Eqs.~(\ref{132})-(\ref{133}) the $\Omega^{\sigma}_{\bfR}$ curvature will insert a Lorentz force 
on a moving through the texture electron leading to transverse current, in complete analogy to the ordinary Hall effect 
in presence of an external magnetic field. It thus makes sense to think of $\bfOmega^{\sigma}_{\bfR}$ as an {\it emergent} magnetic field
$\bfB_e^{\sigma}$. In contrast to the ordinary Hall effect, the Hall effect due to $\bfB_e^{\sigma}$ is opposite
for opposite spins, and it is called the {\it topological Hall effect}. The topological Hall effect is a purely
geometrical phenomenon in that it is driven by (adiabatically slow) modulation of the magnetic texture in space, 
and in that it can be written in terms of Berry curvature of spin-up and spin-down electrons, which sums up to 
zero when both spins are occupied.
 
The magnitude
of the emergent magnetic field at given point $\bfR$ directly depends on how quickly the texture
changes in real space, according to Eq.~(\ref{159}). In case of a texture which exhibits a periodic modulation 
in space, such as~e.g.~in the case of a skyrmion lattice in MnSi, it is useful to define the averaged over 
the magnetic unit cell emergent magnetic field. From the analogy between the real-space Berry curvature
and curvature of the Dirac monopole in $\bfn$-space, we can see that the integral of the emergent
field over the unit cell is proportional to the solid angle which the vector $\bfn$ ``draws" on a sphere
as $\bfR$ is varied in the unit cell. If the texture is such that $\bfn$ covers the sphere completely,
then the integral emergent field is proportional to $4\pi$, and the constant of proportionality is the integer number of times that $\bfn$ winds around
the sphere. This is the so-called {\it winding number} we also mentioned previosly for Chern insulators. 
For example, in case of a skyrmionic lattice in 
MnSi, the winding number is $-1$. Since in MnSi the skryrmion lattice constant is approximately 
165~\AA, the magnitude of the emergent magnetic field $B^{\downarrow}_e$ can be estimated to 
be $-13$~T~\cite{Ritz}.\iffindex{topological Hall effect}\iffindex{emergent magnetic field}\iffindex{skyrmion}  \\

{\it \textbf{Topological Hall effect in terms of a ferromagnetic medium}.}\\
Let us now try to rewrite everything from the point of view of an electron which follows the direction of
the magnetization of the texture but in the reference frame of the magnetization. In adiabatic approximation,
the expectation value of $\bfsigma$ evaluated on the wavefunction $\psi_{\sigma}(\bfR)$ is aligned along $\bfM(\bfR)$.
We can thus apply the transformation $U^{\dagger}(\bfR)$, inverse to the one given by Eq.~(\ref{U-trans}):
$\psi_{\sigma}(\bfR)\longrightarrow U^{\dagger}(\bfR)\psi_{\sigma}(\bfR)=\psi'_{\sigma}(\bfR)$,
which is proportional to $\Ket{\sigma\mathbf{e}_3}$ from (\ref{e3}),~i.e., the spin part of $\psi'_{\sigma}(\bfR)$ is either parallel
or antiparallel to $z$-axis depending on the spin. We want to ask a question: if $\psi_{\sigma}(\bfR)$ solves the Shr\"odinger equation (\ref{SCH}) with $H(\bfR)$ given by (\ref{3.4-1}), what is the Hamiltonian $H'(\bfR)$ for which $\psi'_{\sigma}(\bfR)$
is the solution of the corresponding Schr\"odinger equation? The answer can be simply obtained by substituting the 
 $U(\bfR)\psi'_{\sigma}(\bfR)=\psi_{\sigma}(\bfR)$ into the Schr\"odinger equation for $\psi_{\sigma}(\bfR)$,
 yielding:
 \begin{equation}
 i\frac{\partial\psi'_{\sigma}(t)}{\partial t} = \left[U^{\dagger}(t)H(t)U(t) -iU^{\dagger}(t)\frac{\partial U(t)}{\partial t}\right]\psi'_{\sigma}(t)=H'(t)\psi'_{\sigma}(t).
 \end{equation}
The Hamiltonian $H'$ consists of two parts. The first part is neither $t$ nor $\bfR$ dependent, since
it is obviously given by:
\begin{equation}
H_0=\frac{\bfp^2}{2}- JM\sigma_z,
\end{equation} 
and it provides thus a uniform ferromagnetic background on 
top of which the evolution of $\psi'_{\sigma}$ takes place. The second part of the Hamiltonian $H'$ is the one which leads
to the non-trivial Berry phase. How does it happen? Within the adiabatic approximation, the wavefunction $\psi'_{\sigma}$
can be represented as:
\begin{equation}\label{ferro}
\psi'_{\sigma}(t)=\psi'_{\sigma}(\bfR(t))=c_{\sigma}(t)\Ket{\sigma\mathbf{e}_3}.
\end{equation}
The $c_{\sigma}(t)$ coefficient thus gives us the phase of the time-evolution, 
which consists of the part due to local collinear Hamiltonian $H_0$ (\ref{ferro}),
and of the second part due to the $-iU^{\dagger}\frac{\partial U}{\partial t}$ term of $H'$. This can be seen by substituting the ansatz for $\psi'$ into the corresponding Schr\"odinger equation. Generally speaking, $U$ and $U^{\dagger}$ are matrices in
spin space, which will lead to coupling between the two spin channels. Within the adiabatic
approximation we are going to neglect this coupling, which allows us to consider the scalar quantities $-iU^{\dagger}_{\sigma}\frac{\partial U_{\sigma}}{\partial t}$, given by the diagonal components of $-iU^{\dagger}\frac{\partial U}{\partial t}$, and opposite in
sign for opposite $\sigma$. The contribution of the magnetization chirality to $c_{\sigma}(t)$, which we denote by $c_{\sigma}^U(t)$,
is given by:
\begin{equation}
c_{\sigma}^U(t)=\exp\left( i\int_0^t iU_{\sigma}^{\dagger}(\tau)\frac{\partial U_{\sigma}(\tau)}{\partial\tau}d\tau\right).
\end{equation}
From the latter expression it follows that if we start our evolution from some point $\bfR_0$ at $t=0$ and move our function
to $\bfR(t)$,
which is connected with $\bfR_0$ by path $\calC$, the spin part of the evoluted wavefunction, with the direction of spin being
at all time either parallel or antiparallel to the $z$-axis, can be written as:
\begin{equation}
\psi'_{\sigma}(\bfR) = \exp\left( i\int_0^t iU_{\sigma}^{\dagger}(\tau)\frac{\partial U_{\sigma}(\tau)}{\partial\tau}d\tau\right)\psi'_{\sigma}(\bfR_0)=\exp\left( i\int_{\bfR_0}^{\bfR} \bfcalA^{\sigma}(\bfR)\,d\bfR\right)\psi'_{\sigma}(\bfR_0),
\end{equation}
 which is equivalent to a situation of the AB-effect from section~2.3, with the vector potential given by:
 \begin{equation}
 \bfcalA^{\sigma}(\bfR) = iU_{\sigma}^{\dagger}(\bfR)\nabla_{\bfR} U_{\sigma}(\bfR).
 \end{equation}
 Written explicitly by referring to (\ref{U-trans}) the vector potential reads:
 \begin{equation}\label{gauge-AB}
 \bfcalA^{\sigma}(\bfR) = -\frac{\sigma}{2}\left(1-\cos \theta(\bfR)\right)\nabla\varphi(\bfR),
 \end{equation} 
while its curl gives rise to the effective magnetic field. Both the vector potential given by the latter relation 
and the corresponding
magnetic field are given by exactly the same expressions as the Berry connection and Berry curvature = emergent 
magnetic field from before. Moreover, analogously to the case of AB-effect, given the instantaneous eigenstates of 
$H_0$, $\Ket{\sigma \mathbf{e}_3}_0$,  we can construct the wavefunctions $\Ket{\bfR\sigma \mathbf{e}_3}$ according
to equation (\ref{AB-section}) where $\bfA$ should be replaced with $\bfcalA^{\sigma}$, and this will give us the Berry connection identical to (\ref{gauge-AB}). It can be readily shown,
be referring to the apparatus of section~2.3, that $\Ket{\bfR\sigma \mathbf{e}_3}$ is the instantaneous eigenstate of
the following Hamiltonian:
\begin{equation}
H^{\sigma}_{\rm eff} = \frac{1}{2}\left(\bfp + \bfcalA^{\sigma}\right)^2 - JM\sigma 
\end{equation}
We have thus explicitly shown that the Berry phase problem of Hamiltonian (\ref{3.4-1}) can be also recast as an Aharonov-Bohm
Berry phase problem of an effective ferromagnetic medium subject to a magnetic field, opposite for electrons of opposite spin.
This field is identical to the emergent magnetic field, and gives rise to the topological Hall effect. \\

{\it \textbf{General case}.}\\
While a detailed analysis of the model Hamiltonian given by (\ref{3.4-1}) is extremely insightful, the realistic
Hamiltonian of a solid which exhibits a spatially varying magnetic texture is more complex, in particular,
it includes the crystal potential and spin-orbit interaction. From
the viewpoint of the previous section, the Hamiltonian of our system can be written as $H=H(\bfk,\bfR)$,
where we take the $\bfk$ as the ``local" Bloch vector at point $\bfR$, we exclude explicit dependence
of the Hamiltonian on time (due to~e.g.~moving spin-texture), and we assume that the crystal potential
does not depend on $\bfR$, so that the $\bfR$-dependence appears only due to magnetic texture. For
each band the components of the connection $\calA^n$ and curvature $\Omega^n$ forms in $(\bfR,\bfk)$ space are given by Eqs.~(\ref{potentials}), (\ref{curvatures}) and (\ref{curvature}).

Let us first consider the case without spin-orbit coupling. In this case the spin part and the orbital part
of the wavefunctions are decoupled from each other. For a locally ferromagnetic crystal at point $\bfR$
without spin-orbit interaction $\Omega^n_{\bfk\bfk}\equiv 0$. It is also clear  that without spin-orbit coupling 
$\Omega^n_{\bfk\bfR}=\partial_{\bfR}\calA^n_{\bfk} - \partial_{\bfk}\calA^n_{\bfR}\equiv 0$, since without
spin-orbit $\calA^n_{\bfk}$ does not depend on $\bfR$, and $\calA^n_{\bfR}$ does not depend on $\bfk$.
Thus, the only non-vanishing component of the curvature is the $\Omega^n_{\bfR\bfR}$. In this
case the only effect of the varying spin-texture is the spin-rotation of the wavefunctions for each band and
at each $\bfk$ into the local spin-quantization axis, specified by $\bfR$. Since without spin-orbit coupling rotation
of the spin-quantization axis does not lead to any changes in the spectrum or orbital parts of the wavefunctions, the problem can be effectively re-written in the spin-space, and the apparatus we developed
before can be employed to the full extent. Namely, the $\Omega^n_{\bfR\bfR}$ corresponds to the $\Omega^n_{\bfR}$ from before, and it is {\it exactly} given by Eq.~(\ref{159}).

Without spin-orbit interaction, the relation between the topological and ordinary Hall effects is essentially the same as 
that between the AHE and SHE (without spin-flip SOC): while one is the sum, the other is the difference of 
the intrinsic Hall (ordinary Hall in case of THE and OHE) effects for spin-up and spin-down electrons separately.
Since the texture varies in space slowly, the electrons experience a sufficient amount of disorder-driven momentum-scattering on the scale of a texture, so that the relaxation approximation to the Boltzmann 
equation could be used, leading to the following expression for the topological Hall conductivity (for the
texture in-plane and emergent field out of plane)~\cite{Ritz}:
\begin{equation} 
\sigma_{xy}^{\rm THE} \approx \sum_{\sigma}\sigma|B^{\sigma}_{e}|\sum_n\int_{\rm BZ} 
\tau_{\sigma n}^2 \left(
\frac{\left(v^x_{\bfk n}\right)^2}{m^{yy}_{\bfk n}}   -  \frac{v^x_{\bfk n}v^y_{\bfk n}}{m^{xy}_{\bfk n}}
\right)\frac{\partial f_0(\varepsilon^{\sigma}_{\bfk n})}{\partial \varepsilon}\frac{d^3k}{(2\pi)^3},
\end{equation}
where $m_{\bfk n}$ is the mass tensor, and $\tau_{\sigma n}$ are the relaxation times. The ordinary
Hall conductivity obtained within the same approximations in the orbital magnetic field $|B^{\sigma}_{e}|$
would read the same as the expression above, but without $\sigma$ explicitly in the sum.\\

{\it \textbf{Mixed Berry curvature, Dzyaloshinskii-Moriya interaction 
and charging of skyrmions}}.\\
In the presence of spin-orbit coupling the $\Omega_{\bfk\bfk}^n$ part of the Berry curvature tensor is
different from zero, and this leads to the co-existence of the anomalous Hall and topological Hall contributions
to the transverse conductivity, as seen experimentally. Another manifestation of the spin-orbit
interaction is the fact that the mixed Berry curvature $\Omega^n_{\bfk\bfR}$ is not anymore
vanishing. A profound consequence of this fact is that the modified density of states in $(\bfk,\bfR)$-space, given by (\ref{D-den}), neglecting second and higher order terms, is given by~\cite{Bamler}:
\begin{equation}
D^{n}(\bfk,\bfR) = \frac{1}{(2\pi)^d}\left( 1 - \sum_i \Omega^n_{\bfk\bfR,ii} \right).
\end{equation} 
We can now write again the expression (\ref{FREE}) for the free energy at a certain point $\bfR$, 
as we did for deriving the 
orbital magnetization, but now with a different modified density of states:
\begin{equation}
F (\bfR)= -\frac{1}{\beta(2\pi)^d}\sum_n\int d\bfk\left(1 - \sum_i \Omega^n_{\bfk\bfR,ii}\right)
\ln \left(  1 + e^{-\beta(\mathcal{E}_{n\bfk\bfR}-\mu)} \right),
\end{equation}
where, as previously, $\mathcal{E}_{n\bfk\bfR}$ is given by (\ref{ENERGY}) and contains a correction
$\delta\varepsilon_{n\bfk\bfR}$ due to the texture in addition to the band energy. In the
case of the constant magnetic field we were able to reduce the expression for $\delta\varepsilon_{n\bfk\bfR}$ to $\bfk$-derivatives only. This cannot be done for the spin textures however, since
the dependence of the wavefunctions on $\bfR$ is not only via the $\bfk$-vector:
\begin{equation}
\delta \varepsilon_{n\bfk\bfR}=- {\rm Im}
\Braket{\partial_{\bfR} u_{n\bfk\bfR}|\varepsilon_{n\bfk\bfR} - H_{\bfk\bfR}|\partial_{\bfk} u_{n\bfk\bfR}}
\end{equation}
 We can now
extract the contribution to the free energy which comes from the chirality of the magnetization.
To first order in gradients of the magnetization, it reads:
\begin{equation}\label{dF-DM}
\delta F(\bfR)= \frac{1}{(2\pi)^d}\sum_n\int d\bfk \left( 
f_{\bfk n}\,\delta\varepsilon_{n\bfk\bfR} + \frac{\sum_i \Omega^n_{\bfk\bfR,ii}}{\beta}\ln \left(  1 + e^{-\beta(\varepsilon_{n\bfk\bfR}-\mu)} \right)
\right),
\end{equation} 
where $f_{\bfk n}$ is the Fermi occupation function. The correction due to the chirality
can be expanded in terms of the gradient of the magnetization as:
\begin{equation}
\delta F(\bfR)=D_{ij}(\bfR)\,\hat{\mathbf{e}}_i\cdot\left(   \hat{\bfn} \times \partial_{R_j}\hat{\bfn} \right),
\end{equation}
where $D_{ij}$ correspond to the {\it Dzyaloshinskii-Moriya interaction} (DMI)~\cite{DMI}, which energetically  favors the
chirality of magnetization and which is discussed in detail by Stefan Bl\"ugel in his manuscript C4 of the current book.
\iffindex{Dzyaloshinskii-Moriya interaction} 
From (\ref{dF-DM}) we can derive that 
\begin{equation}
D_{ij}(\bfR)=\frac{1}{(2\pi)^d}\sum_n\int d\bfk \left( f_{\bfk n} A^{ij}_{n \bfk\bfR }  + \frac{B^{ij}_{n\bfk\bfR}}{\beta}
\ln \left(  1 + e^{-\beta(\varepsilon_{n\bfk\bfR}-\mu)} \right)\right),
\end{equation}
with
\begin{eqnarray}
A^{ij}_{n\bfk\bfR} = &-&\mathbf{e}_i\cdot\mathbf{e}_{\varphi}\,{\rm Im}
\braket{\partial_{\theta} u_{n\bfk\bfR}|\varepsilon_{n\bfk\bfR} - H_{\bfk\bfR}|
\partial_{k_j}u_{n\bfk\bfR}} \\
&-&(\mathbf{e}_i\cdot\mathbf{e}_{\theta}/\sin{\theta})\,{\rm Im}
\braket{\partial_{\varphi} u_{n\bfk\bfR}|\varepsilon_{n\bfk\bfR} - H_{\bfk\bfR}|
\partial_{k_j}u_{n\bfk\bfR}},
\end{eqnarray}
and 
\begin{equation}
B^{ij}_{n\bfk\bfR} = -2\mathbf{e}_i\cdot\left[
\mathbf{e}_{\varphi}\,{\rm Im}\braket{\partial_{\theta}u_{n\bfk\bfR}|
\partial_{k_j}u_{n\bfk\bfR}} -
(\mathbf{e}_{\theta}/\sin{\theta})\,{\rm Im}\braket{\partial_{\varphi}u_{n\bfk\bfR}|
\partial_{k_j}u_{n\bfk\bfR}}
\right],
\end{equation}
where $\mathbf{e}_i$ are the cartesian unit vectors, while $\mathbf{e}_{\theta}=\partial\bfn/\partial\theta$ and $\mathbf{e}_{\varphi}=(1/\sin{\theta})\partial\bfn/\partial\varphi$ are the 
unit vectors on the sphere, Fig.~\ref{Sphere}. At zero temperature the DMI can be written as:
\begin{equation}\label{Dij}
D_{ij}(\bfR)=\frac{1}{(2\pi)^d}\sum_n\int d\bfk \, f_{\bfk n}\left[ A^{ij}_{n \bfk\bfR }  - (\varepsilon_{n\bfk\bfR}-\mu)B^{ij}_{n\bfk\bfR}\right].
\end{equation} 
This expression has a one-to-one resemblance to the formula for the orbital magnetization in
ferromagnets,~Eq.~(\ref{M-orb-0T}), which leads to naming $D_{ij}$ the {\it DMI spiralization}~\cite{DMI}.
The quantity $A^{ij}_{n \bfk\bfR}$ is called the {\it twist torque moment} of state $n$ and
it corresponds to the local orbital moment of a wavepacket in case of the orbital magnetiation.
Obviously, the $B^{ij}_{n\bfk\bfR}$ presents the correction to the DMI spiralization due to the
mixed Berry curvature, playing the same role the $\bfk$-space Berry curvature plays for the
orbital magnetization~\cite{DMI}. Besides fundamental importance, the Berry phase expression
(\ref{Dij}) can be used to compute the DMI spiralization from the electronic structure of the
collinear system with the magnetization pointing  along the direction  which corresponds to $\bfR$.
This presents a great simplification as compared to current approaches,
used to calculate the DMI in solids by explicitly incorporating the chirality of the magnetization
into the calculations. For example, in left handed crystal structure of MnSi
the $D_{ij}=-D\,\delta_{ij}$ and the $D$ can be calculated to be $-4.1$~meV~per~\AA~and 
per 8-atom unit cell, in good agreement to experiment~\cite{Bamler}.

Analogously to the case of the change in the Fermi volume due to modified density of states
we calculated before, we can also consider the change in the charge density due to chirality, which comes from the change 
in the energy of the states and modified density of states:
\begin{equation}\label{delta-rho}
\delta\rho(\bfR)=\sum_n\frac{1}{(2\pi)^d}\int d\bfk \left(
\frac{\partial f_{n\bfk}}{\partial \varepsilon}\,\delta \varepsilon_{n\bfk\bfR}
-f_{n\bfk}\sum_i \Omega^n_{\bfk\bfR,ii}.
\right)
\end{equation}
The change of the charge density consists of two terms, of which the first one is the
Fermi surface and the second one is the Fermi see contribution. Formally, the Fermi
see term corresponds to expression (\ref{kubo-occ}) for the derivative of the electric
polarization of an insulator with respect to $\lambda=\bfR$. In the latter case the
integral of this derivative over $\lambda$ would give the change of polarization upon
varying $\lambda$. Similarly, the integral of (\ref{delta-rho}) over the $\bfR$-texture gives
the additional electric charge $\delta\rho$ of the skyrmion due to the Berry phase effect.
There are two important differences between the two situations, however. The first  is
that in insulating skyrmions the integral of the mixed Berry curvature over $\bfk$ and
$\bfR$ spaces vanishes~\cite{Bamler}. The second one is that in metals there is also a Fermi
surface contribution to the charge, which also vanishes for insulating textures. Thus,
to linear order, the charge of insulating skyrmions is zero and the charging effects
in this case occur due to higher-order terms, which can be recast in terms related to the 
Chern numbers. In metals, on the other hand, the charge evaluated according
to (\ref{delta-rho}) will be screened by conduction electrons and its value will be significantly
reduced. So, in MnSi for example, the computed unscreened value of $\delta \rho$ is
about 0.25 electrons, while the screened charge is reduced by orders of magnitude 
with the distribution of $\delta\rho$ over the skyrmion still significant~\cite{Bamler}.

\end{document}